\documentclass[10pt,amsmath,amssymb,twocolumn,superscriptaddress,groupedaddress,nofootinbib,aps,prd,twocolumn,floatfix]{revtex4-2}

\usepackage{graphicx}  
\graphicspath{{./CSE_plots/}}
\usepackage{bm}        
\usepackage{enumitem}
\usepackage{etoolbox}
\usepackage[colorlinks=true]{hyperref}

\newcommand{\comment}[1]{}

\newcommand{\lr}[1]{ \left( #1 \right) }

\newcommand{\vev}[1]{ \langle \, #1 \, \rangle }

\newcommand{\tr}{ {\rm Tr} \, }

\newtoggle{twocolumn}
\toggletrue{twocolumn}

\usepackage[colorlinks=true]{hyperref}

\begin{document}
\sloppy

\title{Numerical Study of the Chiral Separation Effect in Two-Color QCD at Finite Density}

\author{P.~V.~Buividovich}
\email{pavel.buividovich@liverpool.ac.uk}
\affiliation{Department of Mathematical Sciences, University of Liverpool, Liverpool, L69 7ZL, UK}

\author{D.~Smith}
\email{d.smith@gsi.de}
\affiliation{Institut f\"ur Theoretische Physik, Justus-Liebig-Universit\"at, 35392 Giessen, Germany}
\affiliation{Helmholtz Research Academy Hesse for FAIR (HFHF), Campus Giessen, 35392 Giessen, Germany}
\affiliation{Facility for Antiproton and Ion Research in Europe GmbH (FAIR GmbH), 64291 Darmstadt, Germany}

\author{L.~{von Smekal}}
\email{lorenz.smekal@physik.uni-giessen.de}
\affiliation{Institut f\"ur Theoretische Physik, Justus-Liebig-Universit\"at, 35392 Giessen, Germany}
\affiliation{Helmholtz Research Academy Hesse for FAIR (HFHF), Campus Giessen, 35392 Giessen, Germany}

\date{May 26th, 2021}

\begin{abstract}
We study the Chiral Separation Effect (CSE) in finite-density $SU(2)$ lattice gauge theory with dynamical quarks. We find that the CSE is well described by the free quark result in the high-temperature quark-gluon plasma phase. As one enters the confinement regime with broken chiral symmetry at chemical potential smaller than half of the pion mass, the CSE response is gradually suppressed towards low temperatures in comparison to the free quark result. This suppression can be approximately described by assuming that the CSE current is proportional to the charge density, rather than the chemical potential, as suggested in the literature \href{https://arxiv.org/abs/1712.01256}{[Phys.\ Rev.\ D \textbf{97}, 085020 (2018)]}. We also provide an upper bound on the contribution of disconnected fermionic diagrams to the CSE, which is consistent with zero within our statistical errors and small compared to that of the connected diagrams. Our results are obtained mainly in the QCD-like regime of $SU(2)$ gauge theory at low densities, and hence should be at least qualitatively applicable to QCD as well.
\end{abstract}

\maketitle

\section{Introduction}
\label{sec:intro}

\begin{figure}
  \centering
  \includegraphics[width=0.35\textwidth]{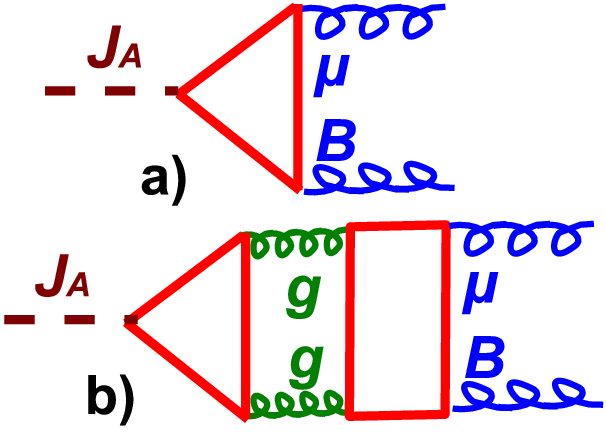}\\
  \caption{a) Feynman diagrams which contribute to the Chiral Separation Effect at leading order and b) one of the possible corrections to it in a gauge theory with dynamical fermions.}
  \label{fig:cse_diagrams}
\end{figure}

Anomalous transport phenomena are transport responses of quantum matter that originate in quantum anomalies, inevitable violations of classical symmetries upon quantization \cite{Landsteiner:1610.04413,Kharzeev:1211.6245}. In particular, in strongly interacting matter described by Quantum Chromodynamics (QCD) the classical symmetry between left-handed and right-handed fermions is violated by the Adler-Bell-Jackiw axial anomaly \cite{Adler:PhysRev69}. This violation manifests itself in the infamous Chiral Magnetic Effect \cite{Kharzeev:0808.3382} - the generation of an electric current along a magnetic field in chirally imbalanced matter - as well as the closely related Chiral Separation Effect (CSE) \cite{Metlitski:hep-ph/0505072,Son:hep-ph/0510049} - the generation of an axial current along an external magnetic field in a dense medium (see Fig.~\ref{fig:cse_diagrams}).

In the last decade, anomalous transport phenomena in QCD matter were systematically and intensely studied in heavy-ion collision experiments at the RHIC \cite{Skokov:1608.00982} and LHC \cite{CMS:1708.01602} colliders, and will also be studied at the NICA collider \cite{NICAWhitePaper}. These studies are not conclusive yet due to large background effects, which contaminate the signatures of anomalous transport \cite{Adam:2006.04251,Lacey:2006.04132,Zhao:1906.11413,Huang:1906.11631,CMS:1708.01602,Bzdak:1207.7327}. A dedicated run with isobar nuclei has been recently completed at RHIC in order to disentangle these background effects \cite{KharzeevNuclearPhysicsNews,Skokov:1608.00982}, and the produced experimental data is currently being analyzed \cite{STAR:1911.00596}.

Just as the viscosity of the quark-gluon plasma is related to hadronic elliptic flow \cite{Teaney:0905.2433}, anomalous transport coefficients characterizing the strengths of the Chiral Magnetic and Chiral Separation Effects can be related to correlations of angular distributions of oppositely charged hadrons in heavy-ion collisions \cite{Voloshin:0806.0029,Upsal:1610.02506}.

One of the most popular ways to interpret experimental data on these correlations relies on the anomalous-viscous fluid dynamics (AVFD) framework \cite{Liao:1611.04586,Liao:1807.05604,Lilleskov:1711.02496}. AVFD is based on anomalous hydrodynamics \cite{Erdmenger:0809.2488,Son:0906.5044,Surowka:0809.2596} which incorporates anomalous transport along with more conventional transport responses such as viscosity and electric conductivity. Hydrodynamic simulation codes which include anomalous transport phenomena on an event-by-event basis are currently being actively developed \cite{Hirono:1704.05375,Liao:1807.05604,Kharzeev:1908.07605,Liao:2004.04440} and are becoming more and more realistic.

The anomalous hydrodynamic description of QCD matter requires the values of anomalous transport coefficients as an input. For a single-component chiral fluid, anomalous transport coefficients are fixed by thermodynamic consistency \cite{Son:0906.5044,Sadofyev:1010.1550}. On the other hand, in a quark-gluon plasma (nearly) chiral quarks interact with dynamical non-Abelian gauge fields, which themselves behave as a viscous fluid and in fact dominate the hydrodynamic flow. When interactions with dynamical gauge fields are present, all anomalous transport coefficients might receive both perturbative \cite{Miransky:1304.4606} (see Fig.~\ref{fig:cse_diagrams}) as well as non-perturbative \cite{Son:hep-ph/0510049,Son:1207.5806,Jensen:1307.3234,Gursoy:1407.3282,Melgar:1404.2434,Avdoshkin:1712.01256} corrections. However, at present not much is known about the magnitude of these corrections.

In a few lattice gauge theory simulations, the Chiral Magnetic \cite{Yamamoto:1105.0385} and Chiral Vortical \cite{Braguta:1401.8095} Effects were studied by measuring the responses of naively discretized axial currents and energy momentum tensors to constant external magnetic or axial magnetic fields (the study \cite{Braguta:1401.8095} of the Chiral Vortical Effect used a trick to replace rotation by a background axial gauge field). In both works \cite{Yamamoto:1105.0385,Braguta:1401.8095} the CME and CVE transport coefficients were found to be 5 to 20 times smaller than those obtained for free quarks, both in high- and in low-temperature phases. If the corrections in the full gauge theory should indeed have the effect to make the anomalous transport responses so small, they might well be unobservable in current heavy-ion collision experiments.

However, the works \cite{Yamamoto:1105.0385,Braguta:1401.8095} used non-chiral lattice fermions with non-conserved currents and an energy-momentum tensor without proper renormalization. On the other hand, a numerical study of the CSE in quenched $SU(3)$ lattice gauge theory with exactly chiral overlap fermions and a properly defined axial current \cite{Buividovich:16:6} found no noticeable corrections to the free quark result in both confinement and deconfinement phases. For free quarks, the axial current induced by the CSE is
\begin{eqnarray}
\label{CSE_free}
 j^A_i  = \frac{\mu \, C_{em}  \, N_c}{2 \pi^2} \, B_i \equiv \sigma_\mathrm{CSE}^0 \, B_i ,
\end{eqnarray}
where $j^A_i = \sum_f \bar{q}_f \gamma_5 \gamma_i q_f$ is the axial current density with quark fields $\bar{q}_f$, $q_f$ of flavor $f$ and $N_c$ colors, $\mu$ is the quark chemical potential, $B_i$ is the magnetic field, and $C_{em} = \sum_f Q_f = \tr\lr{Q}$ is the electromagnetic charge factor in which $Q_f$ denotes the electric charge of quark flavor $f$. The result (\ref{CSE_free}) corresponds to the triangular diagram in Fig.~\ref{fig:cse_diagrams} a). To simplify notation, in what follows we assume that $C_{em}$, which appears in all formulae as a simple pre-factor, is equal to unity: $C_{em} = 1$. The correct value of $C_{em}$ can be restored in all results in an obvious way.

In this paper we study the Chiral Separation Effect in the full gauge theory with dynamical quarks, taking into account the contributions of virtual fermion loops and disconnected fermionic diagrams like the one in Fig.~\ref{fig:cse_diagrams}~b). These contributions are expected to modify the free quark result (\ref{CSE_free}) and are thus important to estimate the detectability of the Chiral Separation Effect in heavy-ion collisions. Rather than studying the theoretically clean, but rather academic, limit of exactly chiral quarks, we address the fate of the CSE in a more realistic setup with finite quark and pion masses and at finite temperatures in the vicinity of the chiral crossover. While giving some general insight into the magnitude of non-perturbative corrections to anomalous transport coefficients, this might also help to estimate the observable consequences of the CSE, such as the electric quadrupole moment of the quark-gluon plasma \cite{Kharzeev:1103.1307} due to Chiral Magnetic Waves \cite{Kharzeev:1012.6026}.

Since the CSE is a feature of finite-density fermions, studying it in QCD would require simulations at finite baryon density, which are complicated by the infamous fermion-sign problem \cite{Gattringer:1603.09517}. With current simulation methods one could only obtain first-principle lattice QCD results for $\mu/T \ll 1$. 

In this work we circumvent the fermion sign problem by using two-color QCD, i.e.\ the $SU(2)$ gauge theory with $N_f=2$ light quark flavours instead of QCD. The path integral weight is manifestly positive in this case, thus the sign problem is absent and the theory can be simulated at finite density. The $SU(2)$ gauge theory is expected to be qualitatively similar to QCD at small densities $\mu < m_{\pi}/2$ \cite{Kogut:hep-lat/0105026,Kogut:hep-ph/0001171}. In this regime there is a conventional QCD-like chiral crossover at some finite temperature, which separates the quark-gluon plasma regime and the hadron gas regime dominated by light pion states \cite{Buividovich:20:1,Hands:1912.10975,Smith:1910.04495,Holicki:1701.04664,Hands:1502.01219,Hands:1210.4496}. We expect that due to this qualitative similarity our numerical study of the CSE in $SU(2)$ gauge theory is also at least qualitatively relevant for real QCD at small densities.

At larger densities, for $\mu > m_{\pi}/2$, $SU(2)$ gauge theory is no longer similar to QCD, since the chiral condensate $\vev{\bar{q} q}$ is rotated into the diquark condensate $\vev{q q}$. Diquarks are bound states of two quarks which are color singlets and hence ``bosonic baryons'' in the $SU(2)$ gauge theory. Instead of the first-order liquid-gas transition of nuclear matter, one therefore observes Bose-Einstein condensation together with a BEC-BCS crossover inside the diquark condensation phase \cite{Smekal:1112.5401,Strodthoff:1306.2897}.  Similarity to QCD, although at a different conceptual level, can be again expected at very large densities and low temperatures, in the conjectured quarkyonic and color-superconducting phases \cite{Pisarski:0706.2191,Braguta:1605.04090}.

\section{Linear response approximation for the Chiral Separation Effect}
\label{sec:CSE_linear}

Within the linear response theory, the Chiral Separation Effect is characterized by the spatial correlator of vector and axial-vector currents $\vev{j^A_i\lr{k} j^V_j\lr{-k}}$, where $k$ is the space-like momentum. Assuming that the only nonzero momentum component is the spatial component $k_3$ \cite{Landsteiner:1102.4577}, at small momenta this correlator can be written as
\begin{eqnarray}
\label{cse_correlator_low_momentum}
 \vev{j^A_1\lr{k_3} j^V_2\lr{-k_3}} = \sigma_\mathrm{CSE} \, k_3 ,
\end{eqnarray}
where $\sigma_\mathrm{CSE}$ is the anomalous transport coefficient in (\ref{CSE_free}) characterizing the strength of the CSE. For free quarks,
\begin{equation}
  \sigma^0_\mathrm{CSE} =  \frac{\mu \,  N_c}{2 \pi^2} \,.
\end{equation}
It is therefore convenient to define a momentum-dependent transport coefficient $\sigma_\mathrm{CSE}\lr{k}$ as
\begin{eqnarray}
\label{cse_correlator_momentum}
 \sigma_\mathrm{CSE}\lr{k_3} \equiv \vev{j^A_1\lr{k_3} j^V_2\lr{-k_3}}/k_3 .
\end{eqnarray}
In the low-momentum hydrodynamic regime, the anomalous transport coefficient $\sigma_\mathrm{CSE}$ in (\ref{CSE_free}) is given by the zero-momentum limit of $\sigma_\mathrm{CSE}\lr{k}$.

For exactly chiral fermions which may interact with each other, but not with other dynamical degrees of freedom (like dynamical gauge fields), $\sigma_\mathrm{CSE}$ is expected to be universal and equal to the free fermion result due to the relation with the Adler-Bell-Jackiw axial anomaly \cite{Metlitski:hep-ph/0505072}. However, corrections are still possible for nonzero quark mass \cite{Son:hep-ph/0510049} and due to fermionic disconnected diagrams like 1b) in Fig.~\ref{fig:cse_diagrams}. As calculations of \cite{Son:hep-ph/0510049} suggest, corrections to the CSE can be related to the amplitude $g_{\pi^0 \gamma \gamma}$ for the $\pi_0 \rightarrow \gamma \gamma$ decay:
\begin{eqnarray}
\label{SonNewmanCSE}
\sigma_\mathrm{CSE} = \frac{\mu N_c \, C_{em}}{2 \pi^2} \lr{1 - g_{\pi^0 \gamma \gamma} + O\lr{\mu}} .
\end{eqnarray}
Within the linear sigma model $g_{\pi^0 \gamma \gamma} = \frac{7 \zeta\lr{3} m^2}{4 \pi^2 T^2}$, where $m$ is the constituent quark mass. In this case, $\sigma_\mathrm{CSE}$ is still approximately linear in $\mu$.

Another calculation of the flavor non-singlet CSE axial current $\vec{j}_A^a$ generated by a finite isospin chemical potential was carried out within chiral effective field theory in \cite{Avdoshkin:1712.01256}. It suggests that in the low-temperature phase, where the CSE current is saturated by pions, $\sigma_\mathrm{CSE}$ is proportional to the isospin charge density $\rho_V^a$ rather than the isospin chemical potential,
\begin{eqnarray}
\label{AvdoshkinCSE}
 \vec{j}_A^a = \frac{N_c \tr\lr{Q}}{\lr{2 \pi \, f_{\pi}}^2} \, \rho_V^a \, \vec{B} .
\end{eqnarray}
While this calculation is not directly applicable to the flavor-singlet axial current in (\ref{CSE_free}) and (\ref{SonNewmanCSE}), at least in the large-$N_c$ limit the flavor-singlet axial current should behave similarly to the flavor non-singlet one \cite{Kaiser:hep-ph/0007101}. Our numerical results presented in Section~\ref{sec:numres} below suggest that a parametrization similar to (\ref{AvdoshkinCSE}) might also work at finite density in the $SU(2)$ gauge theory with dynamical quarks.

\section{Lattice setup}
\label{sec:lattice_setup}

In this work we use the same lattice setup and the same ensembles of gauge field configurations as in our recent works \cite{Buividovich:20:1,Buividovich:21:1}, so here we will provide only a brief summary. We use the standard Hybrid Monte-Carlo algorithm with a tree-level improved Symanzik gauge action and $N_f=2$ flavours of mass-degenerate rooted staggered fermions with bare lattice quark mass $a m_q^{stag} = 5 \cdot 10^{-3}$. This yields a pion mass of $a m_{\pi}^{stag} = 0.158 \pm 0.002$ and the ratio of pion to $\rho$-meson masses $m_{\pi}/m_{\rho} = 0.40 \pm 0.05$.

Our lattices have spatial sizes $L_s = 24$ ($a m_{\pi} L_s = 3.8$) and $L_s = 30$ ($a m_{\pi} L_s = 4.7$) and varying temporal extent $L_t = 4, 6, \ldots, 22$ to control the temperature $T=1/a L_t$. We use a single value of the lattice gauge coupling $\beta = 1.7$, thus working in a fixed-scale approach, which significantly simplifies the analysis of renormalization of lattice observables.

\begin{figure}[h!tpb]
  \centering
  \includegraphics[angle=-90,width=0.5\textwidth]{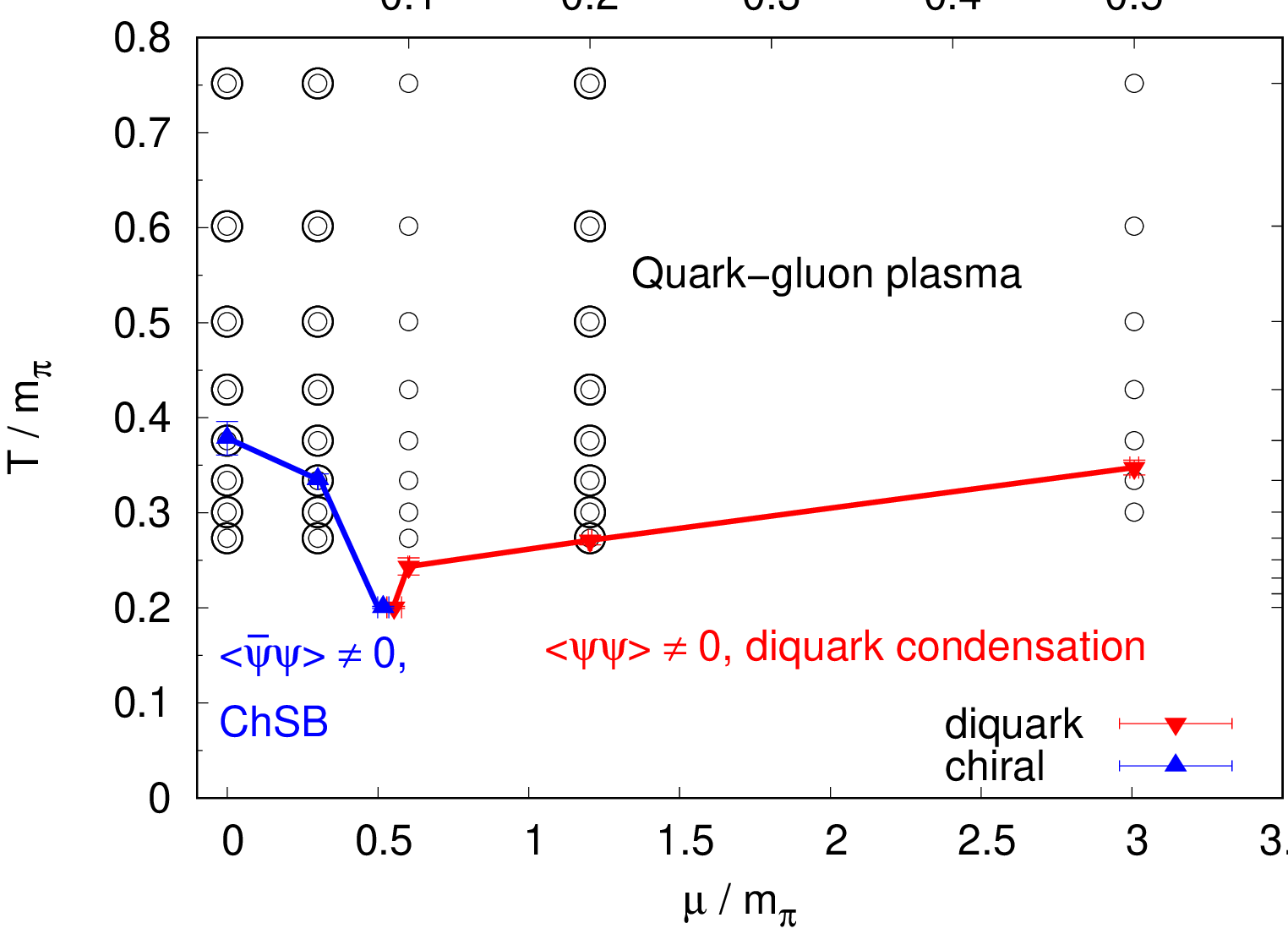}
  \caption{Numerical estimate of the phase diagram of finite-density $SU(2)$ gauge theory with $N_f = 2$ rooted staggered fermions. Blue and red points correspond to inflection points in the $L_t$ dependence of the chiral and diquark condensates, respectively. Configuration sets with lattice size $L_s=24$ only are shown as empty circles, and sets with both $L_s = 24$ and $L_s = 30$ are shown as double circles.}
  \label{fig:phase_diagram}
\end{figure}

Most of our low-temperature ensembles with $L_t \geq 12$ were generated with a small diquark source $\lambda q q$ in the action with $a\lambda = 5 \cdot 10^{-4}$ in order to facilitate diquark condensation, which would otherwise be impossible in a finite volume. This diquark source has very little effect on current-current correlators outside of the diquark condensation phase \cite{Buividovich:20:1}, see also Fig.~\ref{fig:JVJA_vs_lambda}. Estimates of phase boundaries based on our data sets are shown in Fig.~\ref{fig:phase_diagram} (see \cite{Buividovich:20:1} for full details).

We use Domain Wall (DW) and Wilson-Dirac (WD) valence fermions to measure the correlators of axial and vector currents in (\ref{cse_correlator_low_momentum}). On the one hand, for DW fermions the renormalization factor $Z_A$ for the flavour-singlet axial current is expected to deviate from unity by at most few percent \cite{Berkowitz:1704.01114}, which is below our the statistical uncertainty of our Monte-Carlo simulations (and also well below experimental uncertainties). On the other hand, DW fermions are computationally very expensive, and we use the cheaper WD fermions to produce results with better precision covering more points on the phase diagram. A comparison between the results obtained with DW and WD fermions further demonstrates the smallness of axial current renormalization. We do not use staggered valence fermions in order to avoid artifacts related to the unphysical taste symmetry. Such a mixed lattice action with staggered sea fermions and DW valence fermions has already been used in a number of studies of the nucleon axial charge \cite{Edwards:hep-lat/0510062,Berkowitz:1704.01114}.

We tune the bare quark masses $a \, m_q^{DW} = 0.01$ and $a \, m_q^{WD} = -0.21$ in the DW/WD Dirac operators to match the pion mass $am_{\pi}^{stag} = 0.158 \pm 0.002$ obtained with staggered valence quarks. The ratio of pion to rho-meson mass is $m_{\pi}/m_{\rho} \approx 0.4$. To improve the chiral properties of DW and WD fermions without using much finer and larger lattices, we follow \cite{Edwards:hep-lat/0510062} and use HYP smearing \cite{Hasenfratz:hep-lat/0103029} for gauge links in the DW and WD Dirac operators. For DW fermions the lattice size in the fifth dimension is $L_5 = 16$, which is typically sufficient to suppress additive mass renormalization \cite{Edwards:hep-lat/0510062,Berkowitz:1704.01114}.

For WD fermions, we use the conserved vector current
\begin{eqnarray}
\label{lattice_vector_current}
 j^V_{z,\mu} &=& \sum\limits_{x,y} \bar{q}_x \lr{j_{z,\mu}}_{x,y} q_y \,,
 \nonumber \\
 \lr{j_{z,\mu}}_{x,y}
 &\equiv&
 \frac{\partial D_{xy}}{\partial \theta_{z,\mu}}
  \\ &=&
 i P_{\mu}^{+} U_{z,\mu} \delta_{x,z} \delta_{y,z+\hat{\mu}}
 -
 i P_{\mu}^{-} U^{\dag}_{z,\mu} \delta_{x,z+\hat{\mu}} \delta_{y,z}\, ,\nonumber
\end{eqnarray}
where $D_{xy}$ is the Dirac operator,  $x,y,z,\ldots$ label lattice sites, $\gamma_{\mu}$ are the Euclidean gamma-matrices, with $P_{\mu}^{\pm} = (1 \pm \gamma_{\mu})/{2}$, $U_{z,\mu}$ are the $SU(2)$-valued link variables, $\hat{\mu}$ denotes the unit lattice vector in the direction $\mu$, and $\theta_{z,\mu}$ is an external $U(1)$ gauge field. We also use the conventional point-split definition of the axial current for WD fermions \cite{MaianiNPB262},
\begin{eqnarray}
\label{lattice_axial_current}
 \lr{j^A_{z,\mu}}_{x,y}
 = \nonumber \\ =
 i \gamma_{\mu} \gamma_5 U_{z,\mu} \delta_{x,z} \delta_{y,z+\hat{\mu}}
 -
 i \gamma_{\mu} \gamma_5 U^{\dag}_{z,\mu} \delta_{x,z+\hat{\mu}} \delta_{y,z}  .
\end{eqnarray}

For DW fermions, the four-dimensional vector and axial currents are defined in the standard way by summing the five-dimensional conserved current over the fifth dimension. For the vector current a unit weight is used, for the axial current the summation weight changes from $+1$ to $-1$ in the middle of the lattice extending in fifth dimension \cite{Furman:hep-lat/9405004}. The five-dimensional conserved current has a form similar to (\ref{lattice_vector_current}), except that the index $\mu$ takes five values and $x$, $y$, $z$ live on the five-dimensional lattice with open boundary conditions along the fifth dimension.

We measure the contributions of both connected and disconnected fermionic diagrams to the axial-vector current-current correlator in (\ref{cse_correlator_low_momentum}). In coordinate space these contributions are:
\begin{eqnarray}
\label{jAjV_connected}
 \vev{j^A_{x,\mu} j^V_{y,\nu}}_{conn}
 &=&
 \vev{\tr\lr{ j^A_{x,\mu} D^{-1} j^V_{y, \nu} D^{-1} } }\, ,
 \\
\label{jAjV_disconnected}
 \vev{j^A_{x,\mu} j^V_{y,\nu}}_{disc}
 &=&
 \vev{\tr\lr{ j^A_{x,\mu} D^{-1}} \tr\lr{j^V_{y, \nu} D^{-1}} }\, ,
\end{eqnarray}
where the traces are taken over the lattice site, spinor and color indices of the quark fields $\bar{q}$, $q$. The disconnected contribution is measured using standard stochastic estimator techniques. After measuring $\vev{j^A_{x,\mu} j^V_{y,\nu}}_{conn}$ and $\vev{j^A_{x,\mu} j^V_{y,\nu}}_{disc}$ in coordinate space, we perform a discrete Fourier transform to obtain the momentum-space correlators which enter the linear response relations (\ref{cse_correlator_low_momentum}).

\section{Numerical results}
\label{sec:numres}

\begin{figure*}[h!tpb]
  \centering
  \includegraphics[angle=-90,width=0.45\textwidth]{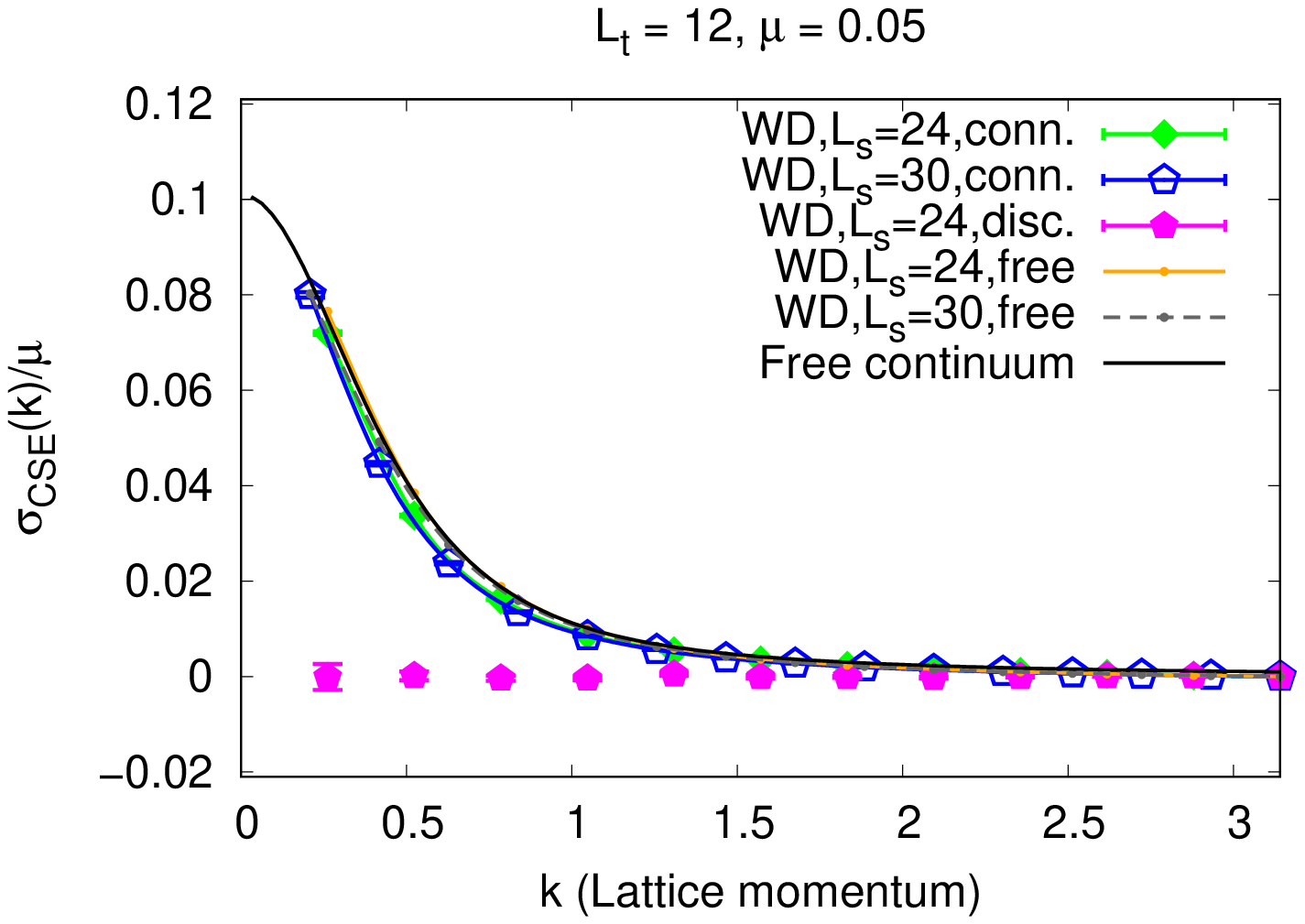}
  \includegraphics[angle=-90,width=0.45\textwidth]{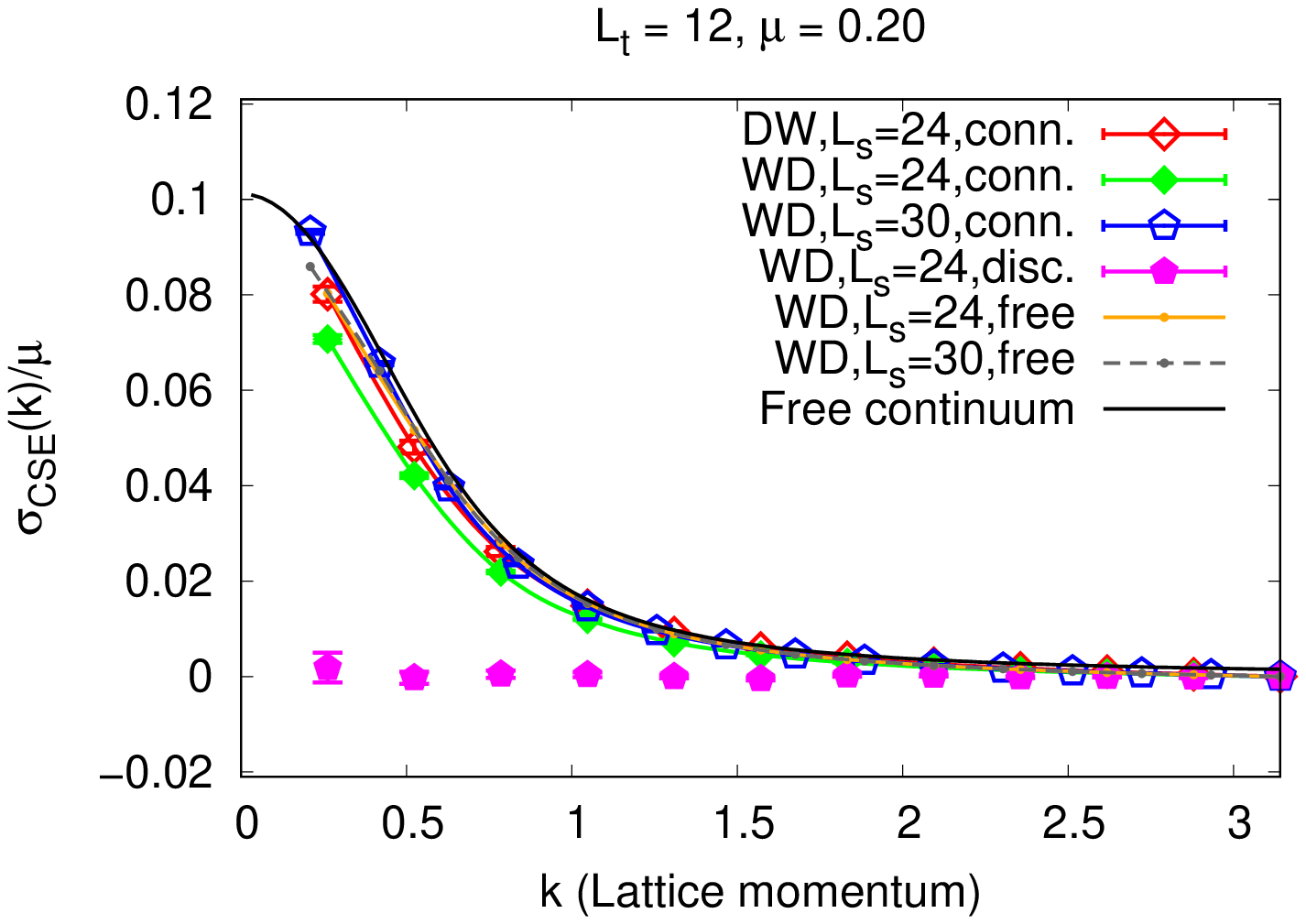}\\
  \includegraphics[angle=-90,width=0.45\textwidth]{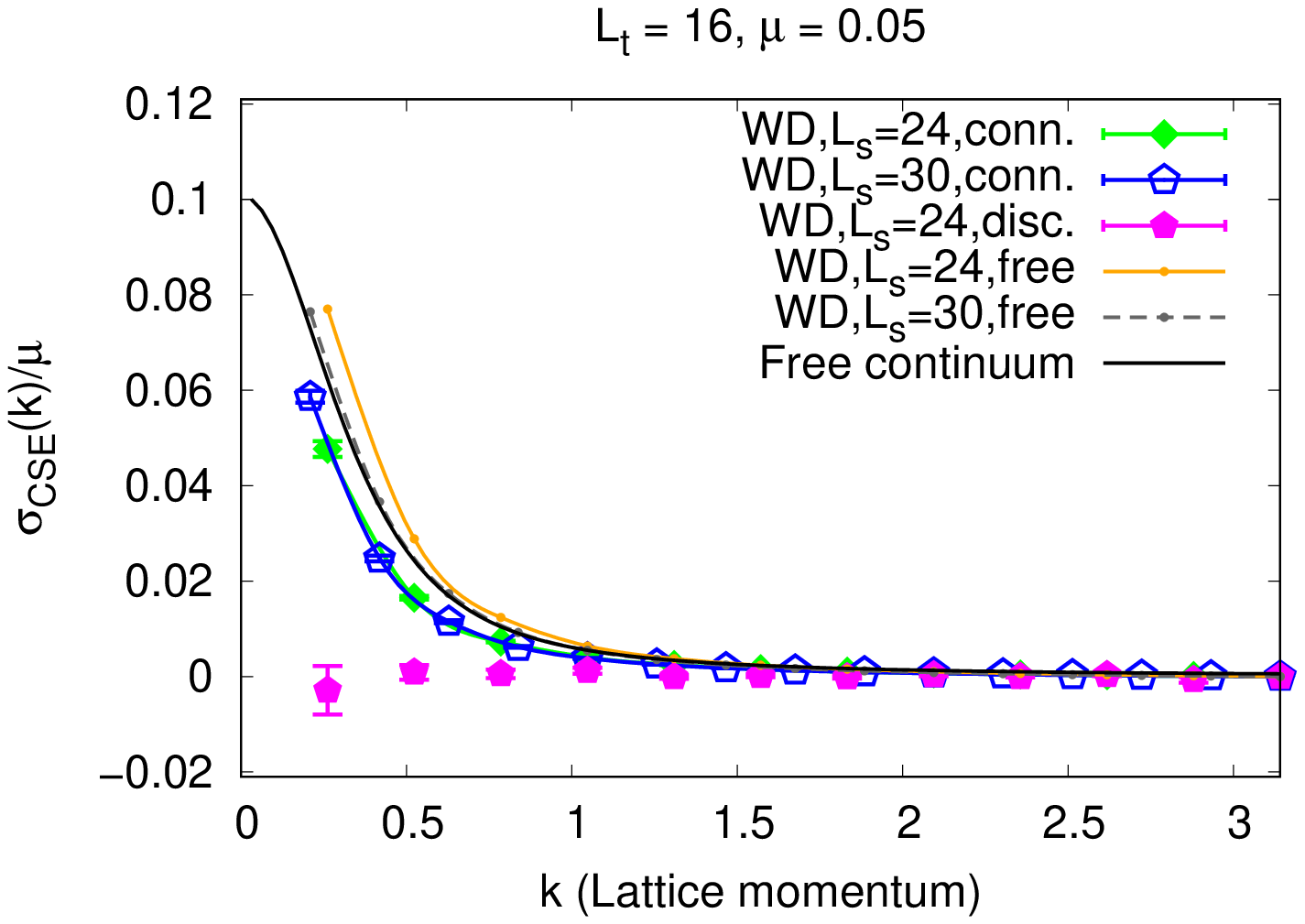}
  \includegraphics[angle=-90,width=0.45\textwidth]{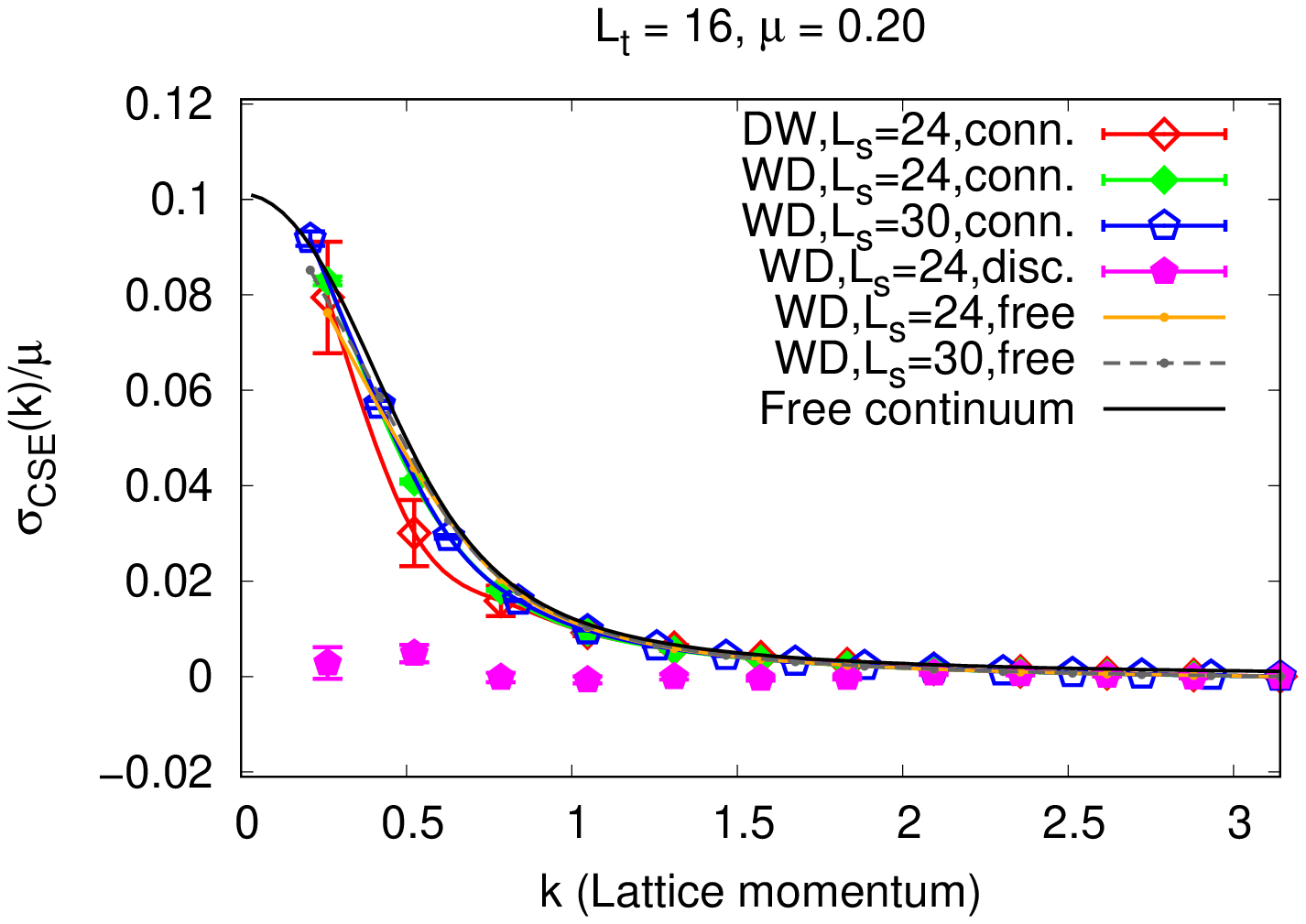}\\
  \includegraphics[angle=-90,width=0.45\textwidth]{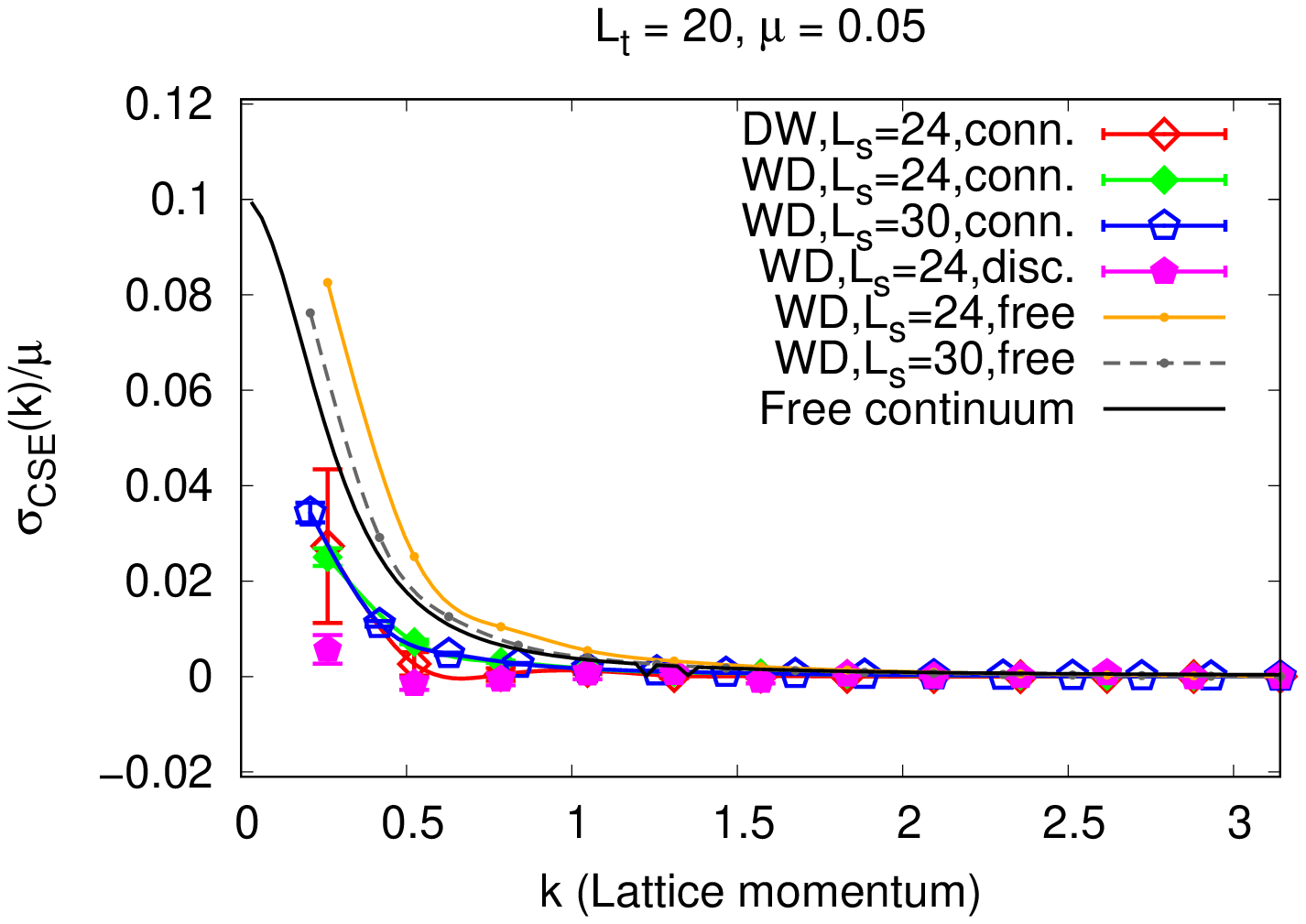}
   \includegraphics[angle=-90,width=0.45\textwidth]{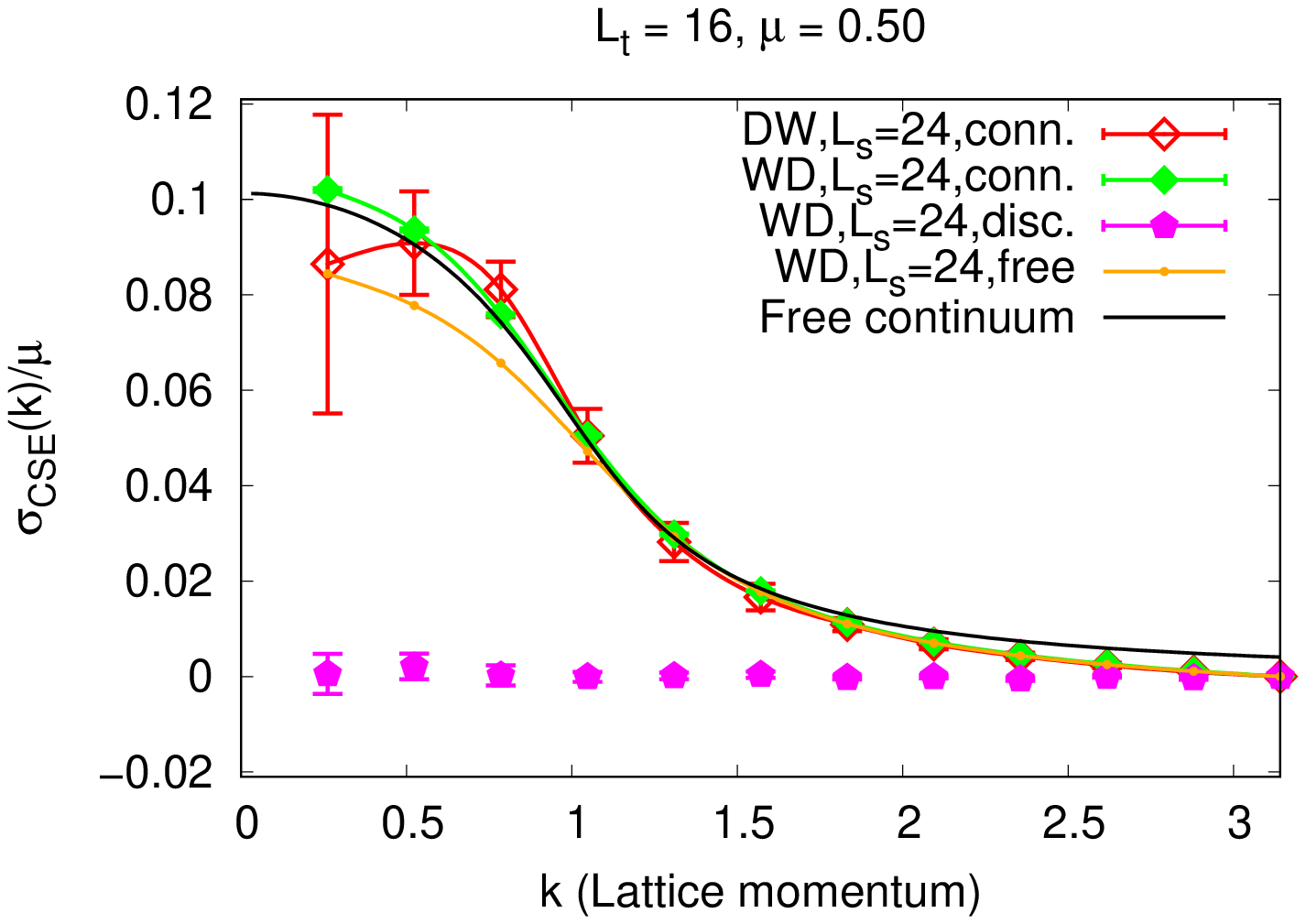}
  \caption{Momentum-dependent CSE transport coefficient $\sigma_\mathrm{CSE}\lr{k}$ as function of lattice momentum $k$ at selected temperatures and chemical potentials, corresponding roughly to: $\mu \simeq 0.32 \; m_\pi$ for three temperatures across the chiral transition (left column), as well as $\mu \simeq 1.3 \; m_\pi $ and $\mu \simeq 3.2 \; m_\pi $ for temperatures approaching the boundary of diquark condensation from above (right column).}
  \label{fig:cse_correlators}
\end{figure*}

In Fig.~\ref{fig:cse_correlators} we present our lattice results for the momentum-dependent CSE transport coefficient $\sigma_\mathrm{CSE}\lr{k}$ defined in (\ref{cse_correlator_momentum}). For comparison, we combine the results obtained with DW and WD fermions, and with the spatial lattice sizes $L_s=24$ and $L_s=30$. For WD fermions on the $L_s=24$ lattices we show the contributions (\ref{jAjV_connected}) and (\ref{jAjV_disconnected}) of both connected and disconnected fermionic diagrams, for other data sets only the connected contributions are shown.

We also compare the gauge theory results with the results obtained for free WD quarks on same lattices. For the free WD quarks, we use a bare quark mass of $am_q^{WD} = 0.01$ (as compared to $am_q^{WD} = -0.21$ in the full gauge theory), since for free quarks there is obviously no mass renormalization. Therefore, in this case we choose the same bare quark mass as for the DW fermions, for which mass renormalization is expected to be weak.

\begin{figure}[h!tpb]
  \centering
  \includegraphics[angle=-90,width=0.45\textwidth]{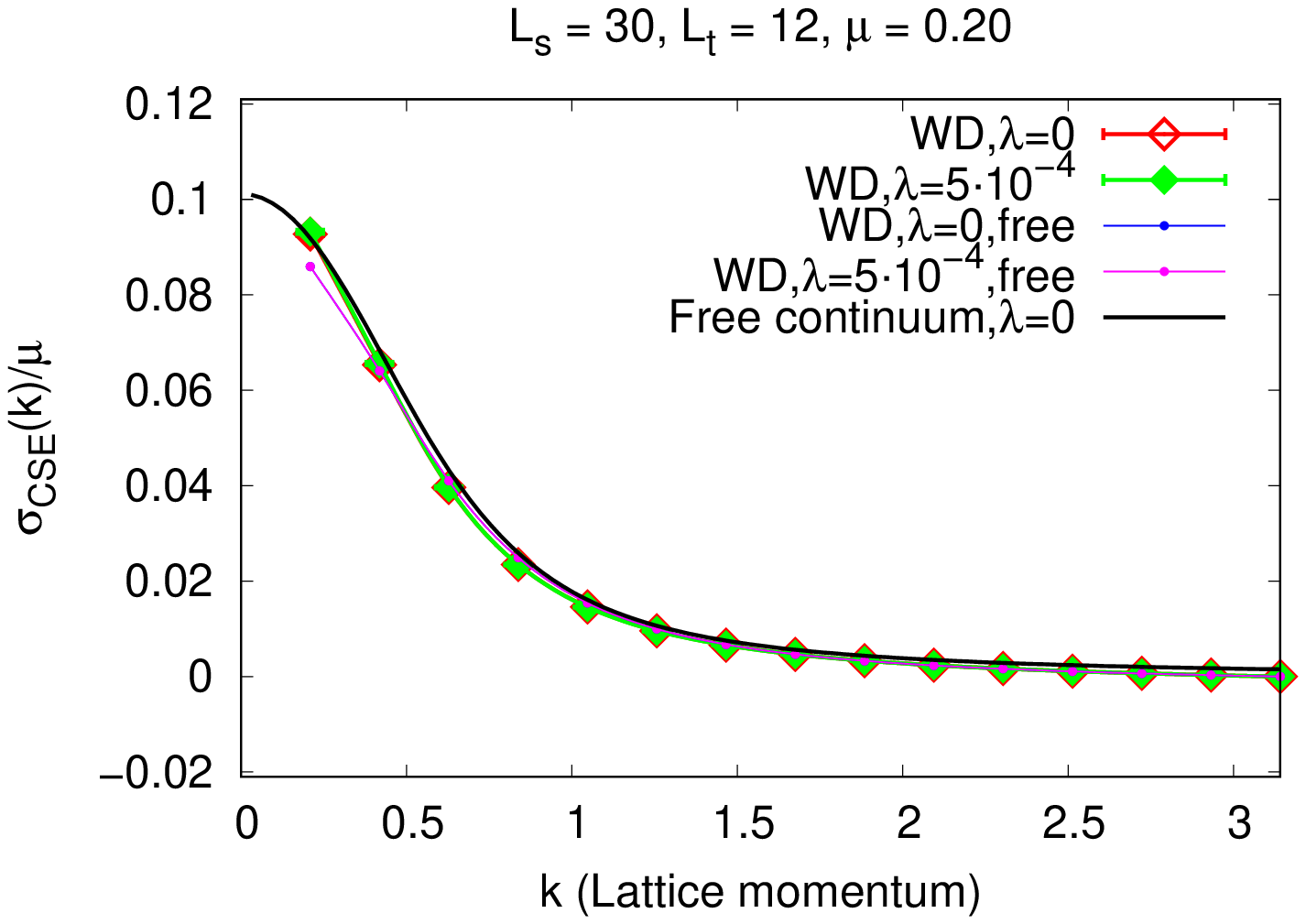}
  \caption{CSE transport coefficient $\sigma_\mathrm{CSE}\lr{k}$ for different values of the diquark source $\lambda$.}
  \label{fig:JVJA_vs_lambda}
\end{figure}

In our calculations we also combine results obtained with zero diquark source $\lambda$ at high temperatures ($L_t < 14$) and with $a\lambda = 5 \cdot 10^{-4}$ at low temperatures ($L_t \geq 14$). In Fig.~\ref{fig:JVJA_vs_lambda} we demonstrate that for these two values of $\lambda$ the CSE transport coefficients $\sigma_\mathrm{CSE}\lr{k}$ are practically indistinguishable.

We see from Fig.~\ref{fig:cse_correlators} that for most values of temperature and chemical potential the momentum-dependent CSE transport coefficient $\sigma_\mathrm{CSE}\lr{k}$ is very close to the corresponding free quark result. An explicit calculation of $\sigma_\mathrm{CSE}\lr{k}$ for free quarks at finite temperature in the continuum is sketched in Appendix~\ref{apdx:cse_free_continuum}. The results of this continuum calculation are shown in all plots of Fig.~\ref{fig:cse_correlators} as solid black lines.

The gauge theory result for $\sigma_\mathrm{CSE}\lr{k}$ only becomes noticeably smaller than the free quark result at small values of the chemical potential $a \mu \lesssim 0.10$ and low temperatures $L_t \gtrsim 16$ (see e.g. the plot for $L_t=20$ and $a\mu=0.05$ corresponding to $\mu = 0.32 \; m_\pi $ in Fig.~\ref{fig:cse_correlators}). In this regime $SU(2)$ gauge theory is expected to be qualitatively similar to real QCD, thus the observed suppression of the CSE in the confined and chirally broken phase is also likely to happen in low-temperature, low-density QCD.

For $a \mu = 0.05$ and $a \mu = 0.20$ we have also calculated $\sigma_\mathrm{CSE}\lr{k \rightarrow 0}$ for two different spatial lattice sizes, $L_s = 24$ and $L_s = 30$. Since the discrete momentum values for both lattices do not coincide, the low-momentum data for both lattices cannot be compared in a direct way. To overcome this difficulty, we construct a third-order spline interpolation of the data for both lattices, which is shown on Fig.~\ref{fig:cse_correlators} as solid lines going through the corresponding data points. Spline interpolations for both lattice sizes coincide with a very good precision for almost all available data sets. We only observe a noticeable difference between the interpolations of the $L_s = 24$ and $L_s = 30$ data on the plot for $L_t = 12$ and $a \mu = 0.20$, with $L_s = 30$ data being much closer to the free quark results. Our data therefore suggest that the suppression of CSE that we observe at low temperatures and densities is not a finite-volume artifact. In fact, the observable finite-volume effects are  somewhat stronger in the high-temperature and high-density regime than at low temperatures.

\begin{figure*}[h!tpb]
  \centering
  \includegraphics[angle=-90,width=0.45\textwidth]{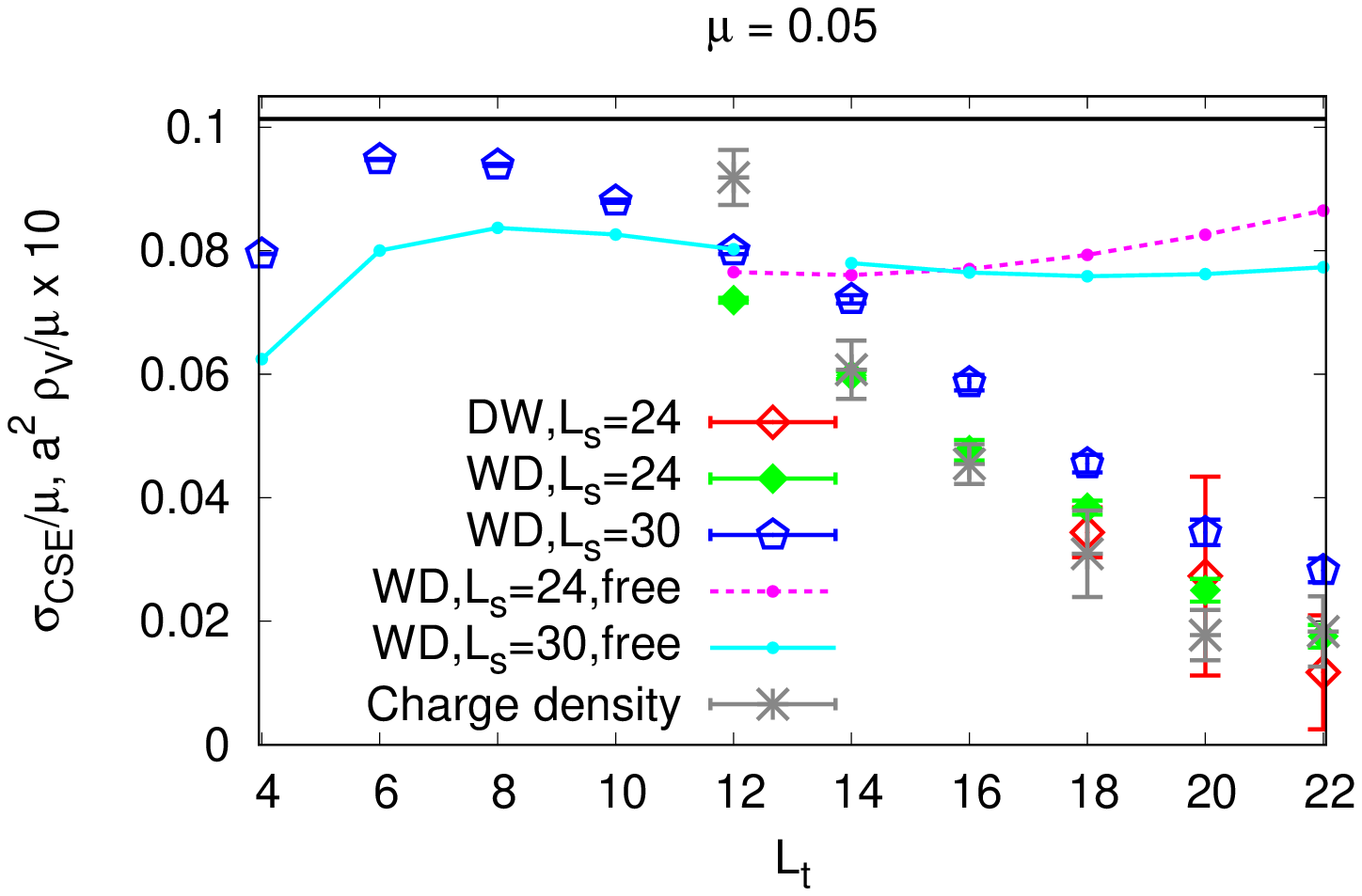}
  \includegraphics[angle=-90,width=0.45\textwidth]{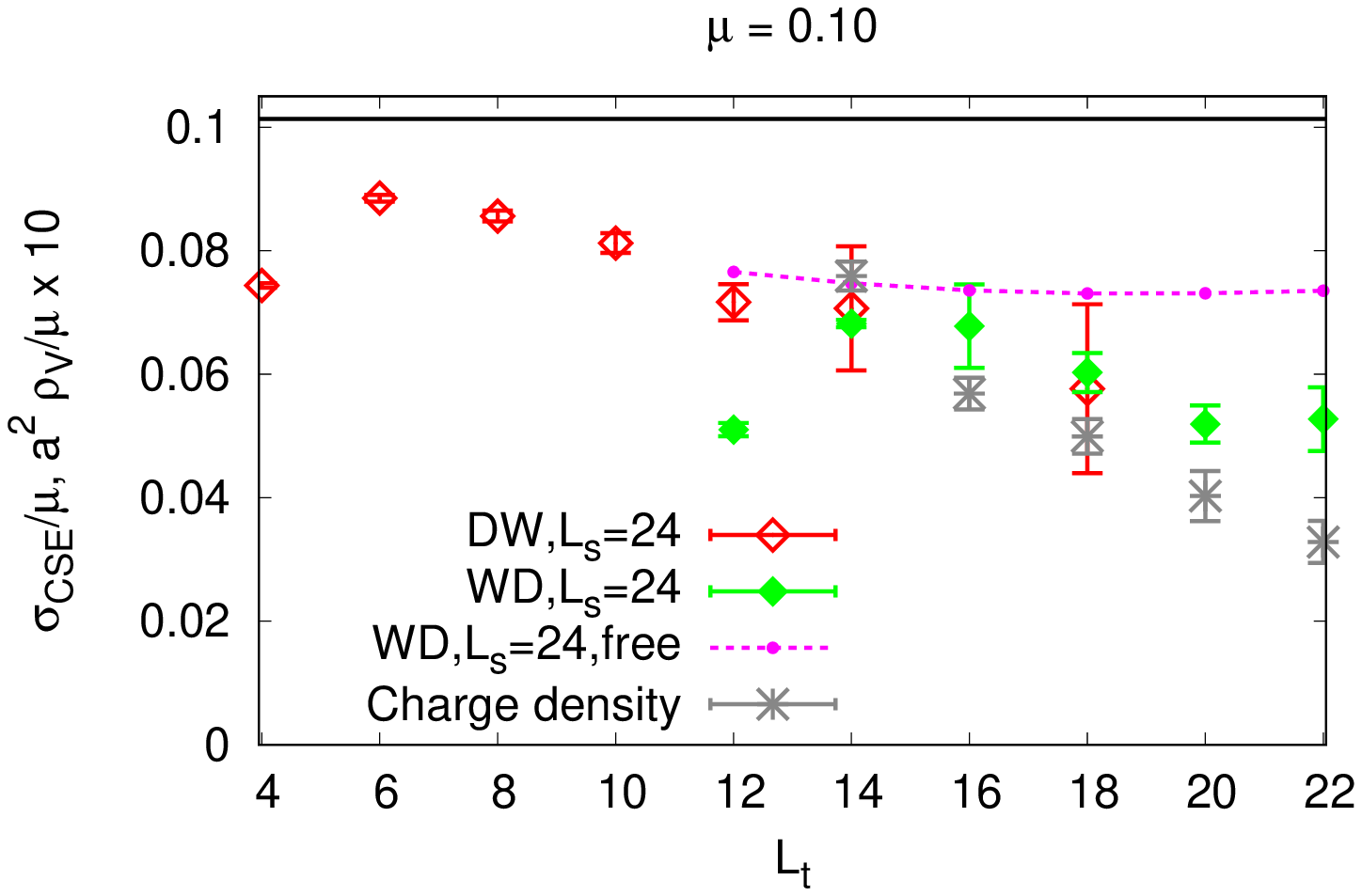}\\
  \includegraphics[angle=-90,width=0.45\textwidth]{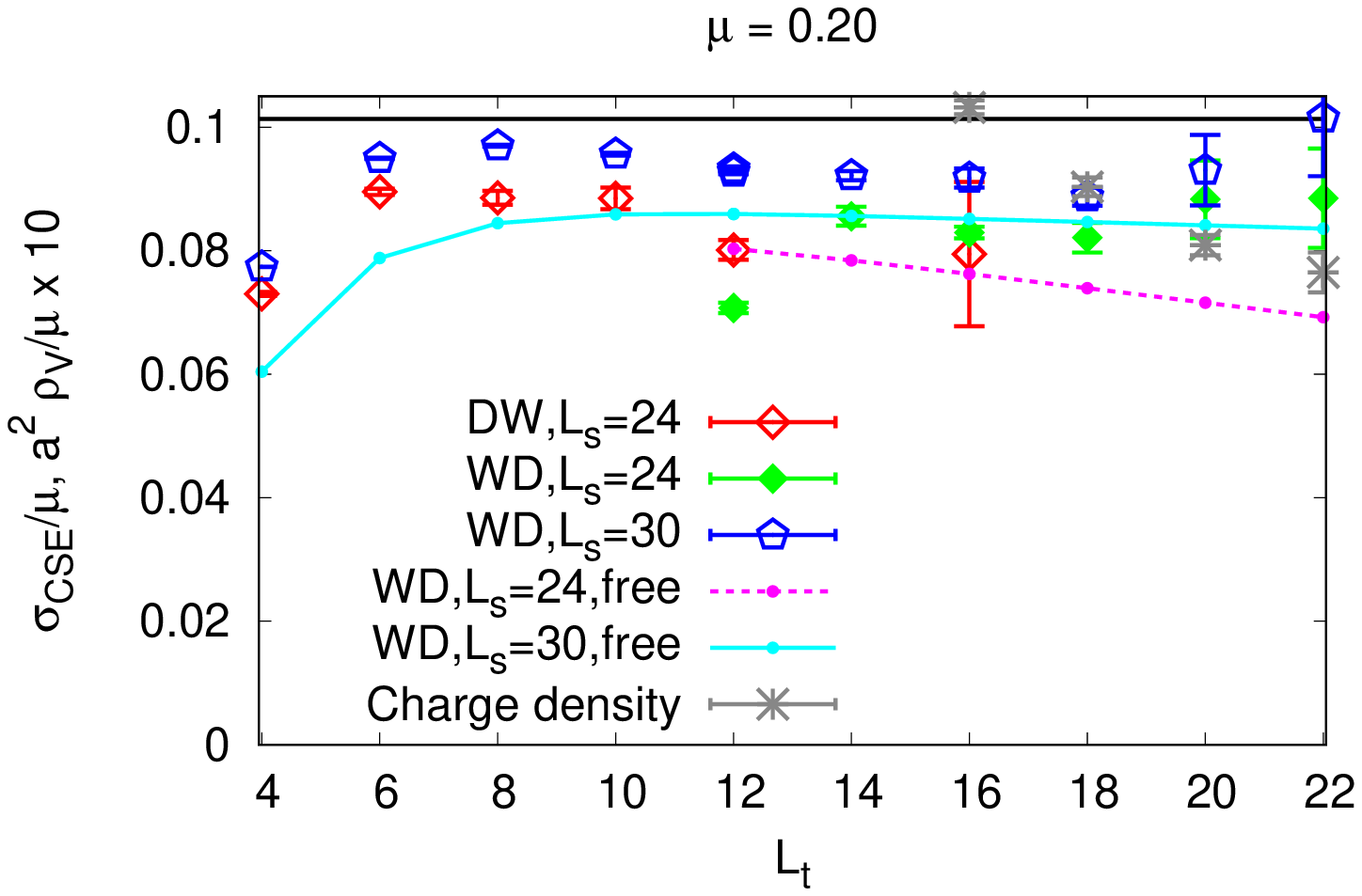}
  \includegraphics[angle=-90,width=0.45\textwidth]{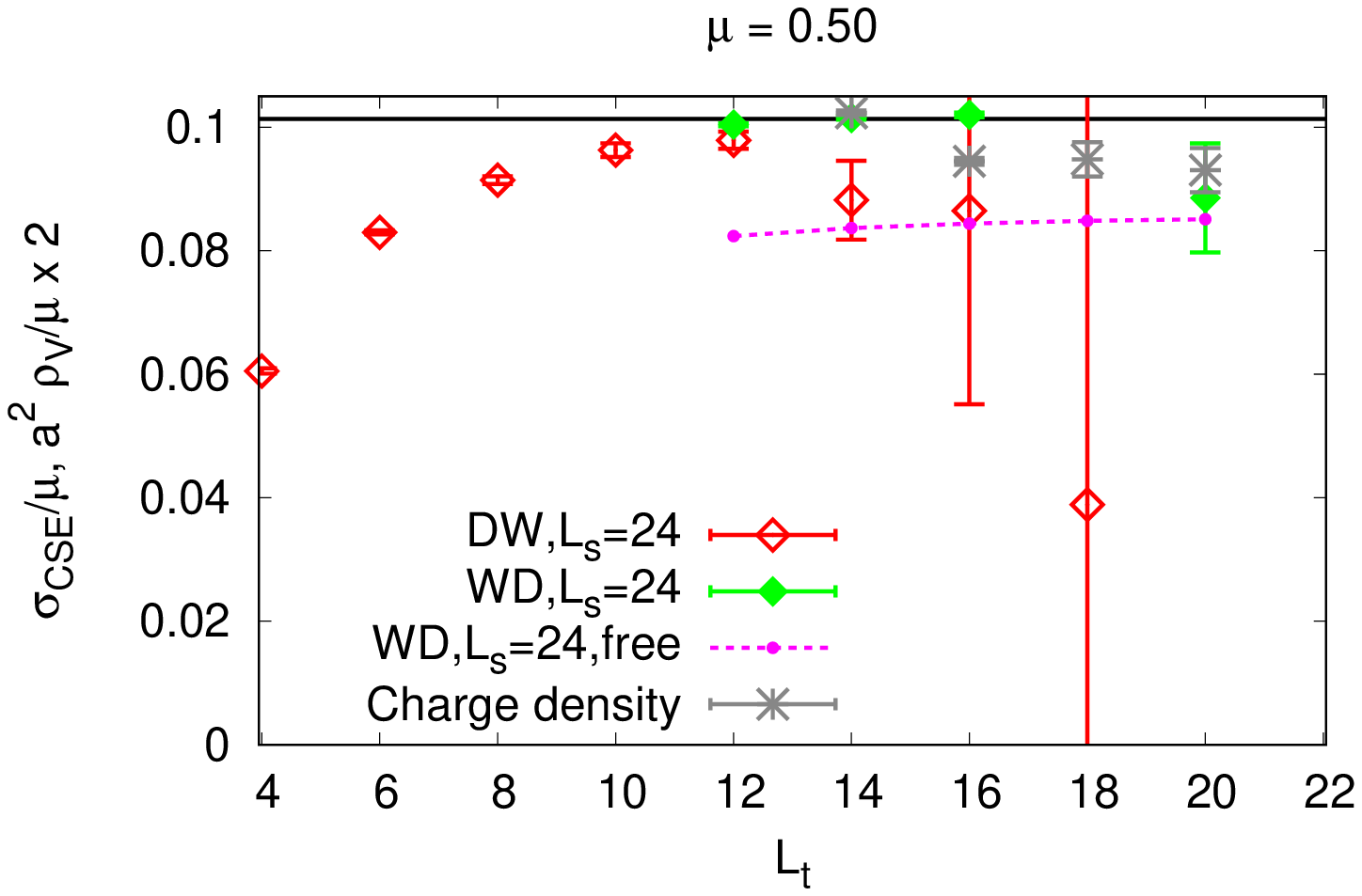}\\
  \caption{Low-momentum limit of the CSE transport coefficient $\sigma_\mathrm{CSE}\lr{k_{min}}$ (where $k_{min} = \frac{2 \pi}{L_s}$ is the smallest nonzero lattice momentum) as a function of temporal lattice size (inverse temperature in lattice units) at different values of the chemical potential. The solid black lines correspond to $\sigma_\mathrm{CSE}\lr{k \rightarrow 0}/\mu =\sigma_\mathrm{CSE}^0/\mu =  {N_c}/(2 \pi^2)$ for free massless quarks in the continuum.}
  \label{fig:cse_summary}
\end{figure*}

The contribution of disconnected fermionic diagrams is consistent with zero within our statistical errors for all values of chemical potential and temperature. The upper bound which we are able to set on these disconnected contributions appears to be least strict for low temperatures and small $\mu$ - that is, exactly in the corner of the phase diagram where also the connected contributions deviate most strongly from the free quark result (see Fig.~\ref{fig:cse_correlators}, plot for $L_t=20$ and $a\mu = 0.05$). We note, however, that the calculation of axial-vector current-current correlators is most difficult precisely in this regime, since because of the small chemical potential the CSE signal is also small compared to statistical fluctuations.

\begin{figure*}[h!tpb]
  \centering
  \includegraphics[angle=-90,width=0.45\textwidth]{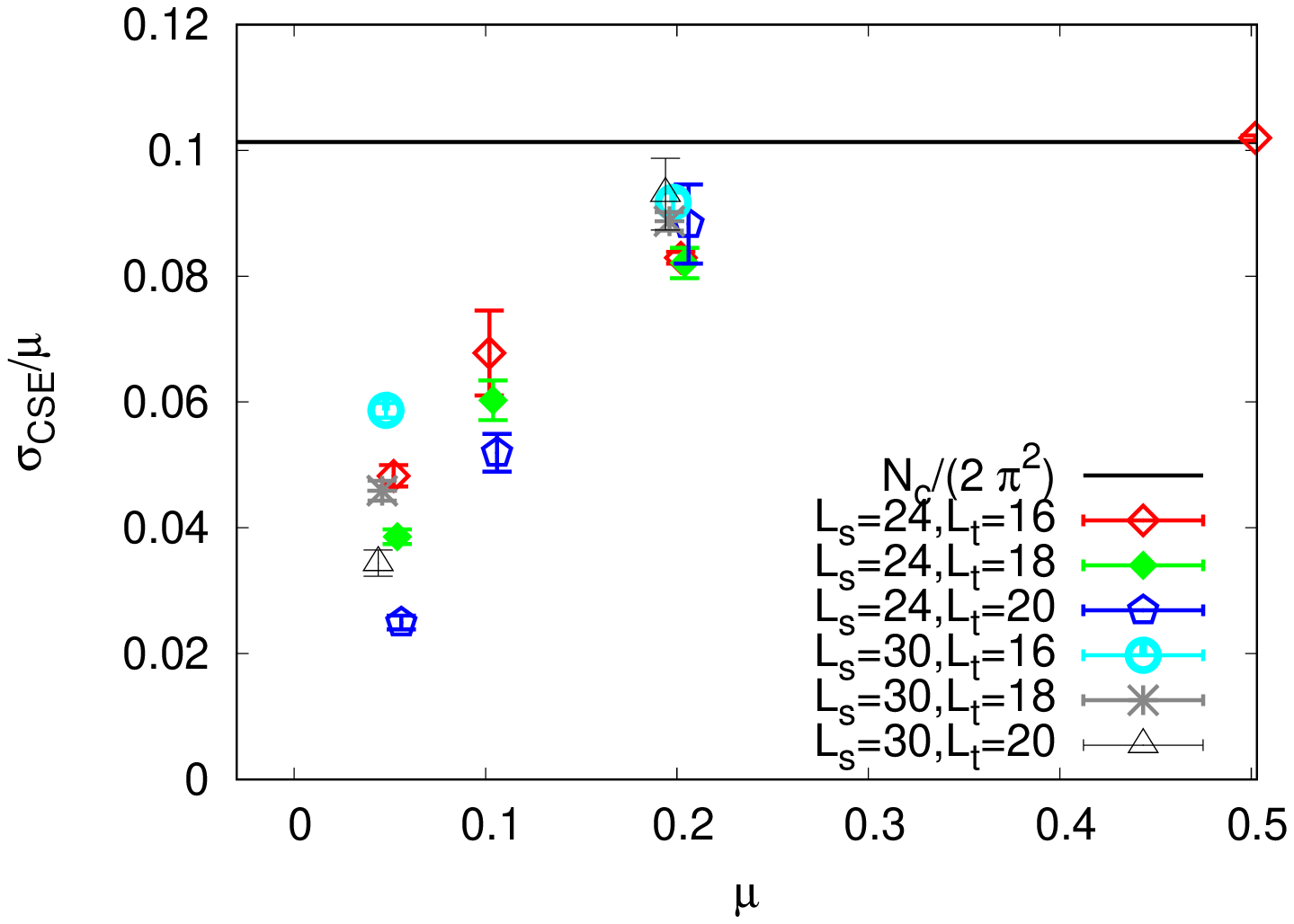}
  \includegraphics[angle=-90,width=0.45\textwidth]{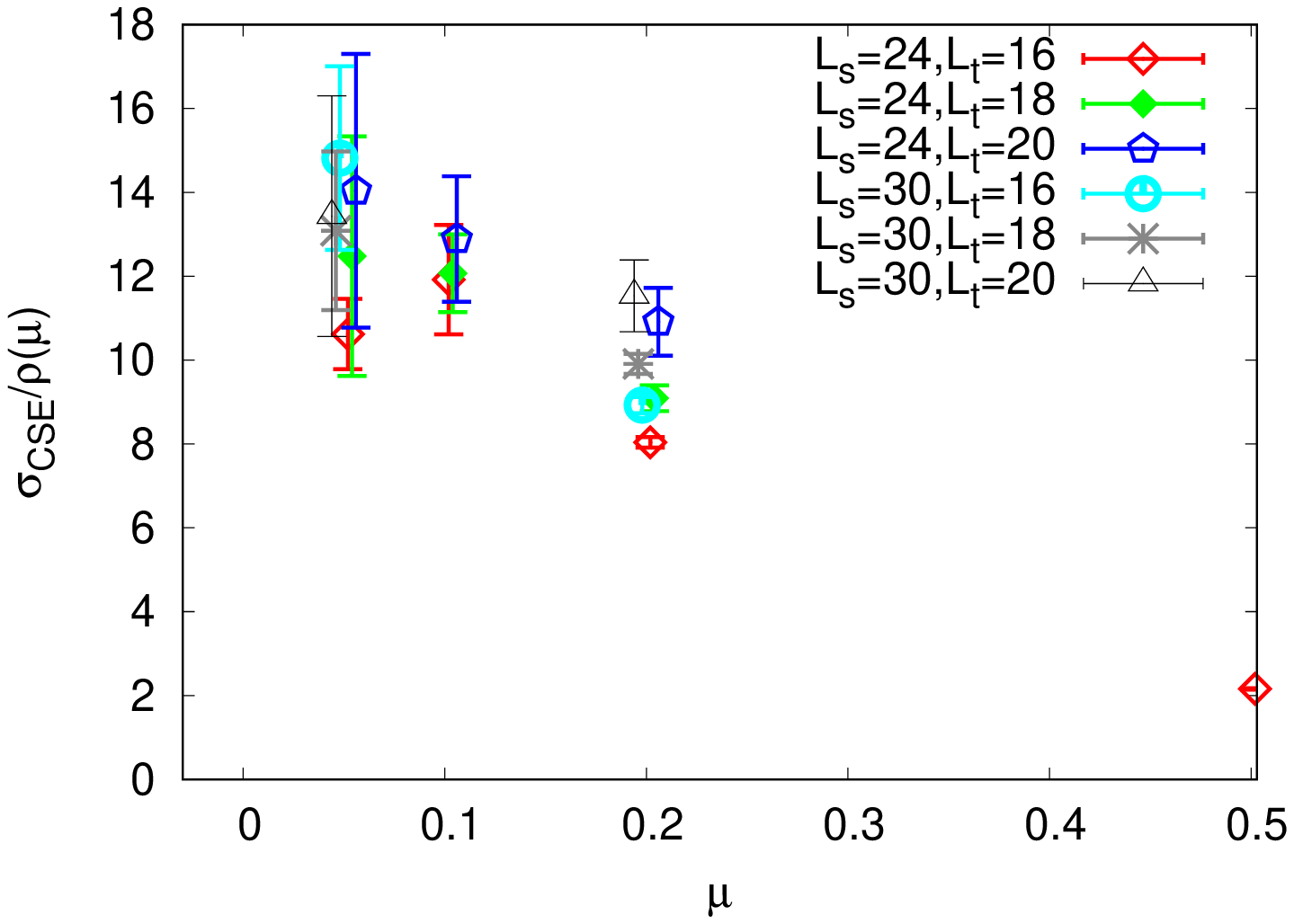}\\
  \caption{The ratio of the low-momentum CSE transport coefficient $\sigma_\mathrm{CSE}\lr{k_{min}}$ and the chemical potential $\mu$ (left), compared to that of $\sigma_\mathrm{CSE}\lr{k_{min}}$  and the charge density $\rho_V(\mu)$ (right). The solid black line in the left plot corresponds to the free continuum quark result $\sigma_\mathrm{CSE}^{0}\lr{k \rightarrow 0}/\mu = {N_c}/(2 \pi^2)$. In the plot on the right, the solid black line marks the optimal value $\alpha = 14 \, a^2$ in our phenomenologically motivated formula (\ref{AvdoshkinCSESinglet}).}
  \label{fig:svsmu}
\end{figure*}

The values of $\sigma_\mathrm{CSE}\lr{k}$ calculated with Wilson-Dirac (WD) and Domain Wall (DW) fermions appear to be very close to each other. This suggests that the effect of multiplicative renormalization of the axial-current operator is small and plays a minor role in comparison with our statistical errors. Indeed, the renormalization factor for the axial singlet current typically appears to be close to unity on fine lattices with sufficiently light pions, especially for Domain Wall fermions \cite{Edwards:hep-lat/0510062,Berkowitz:1704.01114}. For these reasons, we have not determined the precise value of $Z_A$ in this work. Small deviations of $Z_A$ from unity can also be expected to be significantly smaller than any systematic and statistical uncertainties in the experimental detection of anomalous transport phenomena.

Let us now investigate the CSE transport coefficient $\sigma_\mathrm{CSE}\lr{k \rightarrow 0}$ in the low-momentum limit, which is most relevant for anomalous hydrodynamics \cite{Landsteiner:1102.4577}. To this end in Fig.~\ref{fig:cse_summary} we illustrate the temperature dependence of $\sigma_\mathrm{CSE}\lr{k}$ at the smallest nonzero value of the lattice momentum, $a \, k_{min} = 2 \pi/L_s$, and at different values of the chemical potential $\mu$, and compare it with the corresponding results for free quarks on lattices of the same size and for the same momentum.

Again, we observe that $\sigma_\mathrm{CSE}\lr{k_{min}}$ becomes significantly smaller than the free quark result only at low temperatures $L_t \gtrsim 14$ and small values of the chemical potential $\mu \lesssim 0.10$. This region of the phase diagram almost coincides with the QCD-like regime with spontaneously broken chiral symmetry, which should be dominated by pions. In this regime, the free quark result for $\sigma_\mathrm{CSE}\lr{k_{min}}$ has a rather weak temperature dependence, but the gauge theory result is strongly suppressed towards lower temperatures.

A comparison of the values of $\sigma_\mathrm{CSE}\lr{k_{min}}$ for $L_s = 24$ and $L_s = 30$ might make an impression that the suppression of $\sigma_\mathrm{CSE}\lr{k_{min}}$ becomes weaker for larger volumes. However, the apparent deviation of the results for $L_s = 24$ and $L_s = 30$ is caused simply by the difference of the minimal nonzero momenta $a \, k_{min} = \frac{2 \pi}{L_s}$ for different lattice sizes. As one can see from Fig.~\ref{fig:cse_correlators}, the data interpolated to continuum momentum values does not show significant volume dependence.

Let us now try to describe the observed CSE suppression at low temperatures and densities in terms of some phenomenological formula. Let us try a formula of the form (\ref{AvdoshkinCSE}), but with a flavour-singlet current and chemical potential:
\begin{eqnarray}
\label{AvdoshkinCSESinglet}
  \sigma_\mathrm{CSE}\lr{k \rightarrow 0, \mu} &=& \alpha \, \rho_V\lr{\mu},
  \nonumber \\
  \vec{j}_A &=& \alpha \, \rho_V\lr{\mu} \, \vec{B} .
\end{eqnarray}
To this end, in Fig.~\ref{fig:cse_summary} we also show the rescaled charge density $\alpha \, \rho_V\lr{\mu}/\mu$, tuning the coefficient $\alpha$ to achieve the best coincidence between the data points for $\sigma_\mathrm{CSE}\lr{k_{min}}/\mu$ and $\alpha \, \rho_V\lr{\mu}/\mu$. To achieve this we minimize the mean squared deviation of $\alpha \, \rho_V\lr{\mu}/\mu$ from $\sigma_\mathrm{CSE}\lr{k_{min}}/\mu$ at $a \, \mu = 0.05$ for the lattice size $L_s = 30$. We find that at low temperatures $L_t \gtrsim 14$ the dependence of $\sigma_\mathrm{CSE}\lr{k_{min}}/\mu$ for $a \, \mu = 0.05$ on $L_t$ can be indeed well described by the formula (\ref{AvdoshkinCSESinglet}) with $\alpha \approx 14 \, a^2$ (here we simply express $\alpha$ in lattice units, without implying that $\alpha$ scales as $a^2$). The same value of $\alpha$ also describes the data for $a \, \mu = 0.10$ and $L_t \gtrsim 16$ reasonably well.

For larger values of the chemical potential, $a \, \mu = 0.20$ and $a \, \mu = 0.50$, the formula (\ref{AvdoshkinCSESinglet}) does not seem to work. Interestingly, Fig.~\ref{fig:cse_summary} suggests that for these values of $\mu$ the CSE coefficient $\sigma_\textrm{CSE}(k_{min})$ appear to be even slightly larger than the free quark result on the same lattice, although the deviations do not exceed $20 \%$. According to Fig.~\ref{fig:phase_diagram}, the data points for $a \, \mu = 0.20$ and $a \, \mu = 0.50$ are within the quark-gluon plasma regime or at the boundary of the diquark condensation phase, except for the point with $a \mu = 0.50$ and $L_t = 20$ that is in the diquark condensation regime. While one can see a drop of $\sigma_\textrm{CSE}(k_{min})/\mu$ by around $20 \%$ for this single data point, this change is comparable to statistical errors, and we cannot make definite conclusions on the behavior of $\sigma_\textrm{CSE}(k_{min})$ in the diquark condensation phase. We can only rule out a significant suppression of the CSE around the boundary of this regime.

According to the formula (\ref{AvdoshkinCSE}) with $C_{em} = \tr\lr{Q} = 1$, the coefficient $\alpha$ in (\ref{AvdoshkinCSESinglet}) should be related to the pion decay constant $f_{\pi}$ via $\alpha = {N_c}/(2 \pi f_{\pi})^2 $. Using our estimate $\alpha \approx 14 \, a^2$, from this relation we roughly estimate $a \, f_{\pi} \approx 0.06$, which has a reasonable order of magnitude when compared to the mass of the $\rho$ meson,
$a \, m_{\rho} \approx a \, m_{\pi}/0.4 \approx 0.4$ in our calculations. Thus fitting our data with the formula (\ref{AvdoshkinCSESinglet}) implies the ratio ${f_{\pi}}/{m_{\rho}} \approx 0.15$, as compared to ${f_{\pi}}/{m_{\rho}} \approx  0.12 $ with  $f_{\pi} \approx 93 \, \textrm{MeV}$, $m_{\rho} \approx 770 \, \textrm{MeV}$ in QCD.

In order to present further evidence for the scaling of $\sigma_\mathrm{CSE}$ with $\rho_V$, in Fig.~\ref{fig:svsmu} we show the dependence of the ratios $\sigma_\mathrm{CSE}\lr{k_{min}}/\mu$ (on the left) and $\sigma_{CSE}\lr{k_{min}}/\lr{a^2 \rho_V\lr{\mu}}$ (on the right) on the chemical potential $\mu$ at different temperatures with $L_t \geq 16$. It is quite obvious from these figures that the ratio $\sigma_{CSE}\lr{k_{min}}/\lr{a^2 \rho_V\lr{\mu}}$ shows smaller relative deviations from a constant value $\alpha \, a^{-2} = 14$ than the ratio $\sigma_\mathrm{CSE}\lr{k_{min}}/\mu $ does from ${N_c}/(2 \pi^2)$. In particular, for $L_s = 30$ all data points contain the value $\alpha \, a^{-2} = 14$ within their error bars. The data points for other ensembles deviate from this value by not more than $50 \%$, whereas for the ratio $\sigma_\mathrm{CSE}\lr{k_{min}}/\mu$ the relative deviation between different data points is much larger, up to a factor of $5$. Note that the ratio $\sigma_{CSE}\lr{k_{min}}/\lr{a^2 \rho_V\lr{\mu}}$ has larger error bars than $\sigma_\mathrm{CSE}\lr{k_{min}}/\mu$ because $\rho_V\lr{\mu}$ also has statistical errors, in contrast to $\mu$.

While these observations give some qualitative support to the formula (\ref{AvdoshkinCSESinglet}), because of a conceptually different status of axial-singlet and axial-non-singlet currents in low-energy chiral effective theory, at the quantitative level one can expect further corrections to this formula. A general conclusion that we can make based on our results is that the CSE becomes more and more suppressed for low temperatures in the confinement regime and for values of the chemical potential roughly smaller than half of the pion mass. At higher temperatures and densities, the CSE transport coefficient approaches its value for free quarks.

\section{Conclusions}
\label{sec:conclusions}

To summarize, we have found that in a gauge theory with dynamical quarks the Chiral Separation Effect is very close to the free quark result in the high-temperature quark-gluon plasma regime, and is gradually suppressed towards lower temperatures and densities in the low-temperature hadron resonance-gas regime with broken chiral symmetry at $T \lesssim T_c$ and $\mu \lesssim m_{\pi}/2$. Exactly this regime of $SU(2)$ gauge theory is similar to the low-temperature, low-density phase of real finite-density QCD, thus our findings should be also relevant for real QCD at least qualitatively.

Note that our conclusions on the CSE suppression do not contradict the results of a previous study \cite{Buividovich:16:6} in quenched $SU(3)$ lattice gauge theory, where exactly chiral valence quarks were used, thus the pion mass was formally zero and the region with $\mu \lesssim m_{\pi}/2$ was absent.

The suppression of the CSE at low densities and temperatures can be approximately described if one assumes that the CSE current is proportional to the charge density rather than chemical potential, as in equation (\ref{AvdoshkinCSE}). While the formula (\ref{AvdoshkinCSE}) was derived in \cite{Avdoshkin:1712.01256} for the axial non-singlet current, we see that at the qualitative level it also applies to the axial singlet current, which has a different status within the chiral effective theory.

Contributions from disconnected fermionic diagrams to the Chiral Separation Effect appear to be consistent with zero within our statistical errors. The latter are relatively large at low temperatures and densities. For this reason we cannot rule out that the disconnected contribution might become important when the connected one is strongly suppressed. This scenario would certainly be interesting from a theoretical viewpoint. However, as the sum of both contributions still appears to be small when compared to the free massless fermion result, it is probably not very relevant for the experimental detection of anomalous transport phenomena.

One potential source of systematic errors in our study is the multiplicative renormalization of the axial current. However, comparison of the data obtained with Domain Wall and Wilson-Dirac fermions, as well as previous lattice studies of axial current renormalization, suggest that the renormalization effects are small.

Another potential improvement of our study would be the use of twisted boundary conditions in one of the spatial directions in order to access $\sigma_\mathrm{CSE}\lr{k}$ for nonzero momenta $k$ that are smaller than $k_{min} = \frac{2 \pi}{L_s}$ and perform a proper extrapolation of the numerical data for $\sigma_\mathrm{CSE}\lr{k}$ to the limit $k \rightarrow 0$. Fig.~\ref{fig:cse_correlators} suggests that extrapolation from $k_{min}$ towards $k = 0$ can potentially make the estimates of the limit $\sigma_\mathrm{CSE}\lr{k \rightarrow 0}$ considerably larger. To avoid this uncertainty of extrapolation to $k = 0$, in the present work we have compared our data with the free quark result for $\sigma_\mathrm{CSE}\lr{k}$ at the same nonzero momentum $k_{min}$, but comparison at smaller values of $k$ would be more reliable.

\begin{acknowledgments}
The work of P.~B.\ was supported by the Heisenberg Fellowship from the German Research Foundation, project BU2626/3-1. D.~S.\ received funding from the European Union's Horizon 2020 research and innovation programme under grant agreement No.~871072, also known as CREMLINplus (Connecting Russian and European Measures for Large-scale Research Infrastructures).

This work was performed using the Cambridge Service for Data Driven Discovery (CSD3), part of which is operated by the University of Cambridge Research Computing on behalf of the STFC DiRAC HPC Facility (www.dirac.ac.uk). The DiRAC component of CSD3 was funded by BEIS capital funding via STFC capital grants ST/P002307/1 and ST/R002452/1 and STFC operations grant ST/R00689X/1. DiRAC is part of the National e-Infrastructure.

The simulations were also performed on the GPU cluster at the Institute for Theoretical Physics at Giessen University. Many thanks to Dominik Schweitzer for keeping this cluster alive during \#JLUoffline.
\end{acknowledgments}


\begin{thebibliography}{62}%
\makeatletter
\providecommand \@ifxundefined [1]{%
 \@ifx{#1\undefined}
}%
\providecommand \@ifnum [1]{%
 \ifnum #1\expandafter \@firstoftwo
 \else \expandafter \@secondoftwo
 \fi
}%
\providecommand \@ifx [1]{%
 \ifx #1\expandafter \@firstoftwo
 \else \expandafter \@secondoftwo
 \fi
}%
\providecommand \natexlab [1]{#1}%
\providecommand \enquote  [1]{``#1''}%
\providecommand \bibnamefont  [1]{#1}%
\providecommand \bibfnamefont [1]{#1}%
\providecommand \citenamefont [1]{#1}%
\providecommand \href@noop [0]{\@secondoftwo}%
\providecommand \href [0]{\begingroup \@sanitize@url \@href}%
\providecommand \@href[1]{\@@startlink{#1}\@@href}%
\providecommand \@@href[1]{\endgroup#1\@@endlink}%
\providecommand \@sanitize@url [0]{\catcode `\\12\catcode `\$12\catcode
  `\&12\catcode `\#12\catcode `\^12\catcode `\_12\catcode `\%12\relax}%
\providecommand \@@startlink[1]{}%
\providecommand \@@endlink[0]{}%
\providecommand \url  [0]{\begingroup\@sanitize@url \@url }%
\providecommand \@url [1]{\endgroup\@href {#1}{\urlprefix }}%
\providecommand \urlprefix  [0]{URL }%
\providecommand \Eprint [0]{\href }%
\providecommand \doibase [0]{https://doi.org/}%
\providecommand \selectlanguage [0]{\@gobble}%
\providecommand \bibinfo  [0]{\@secondoftwo}%
\providecommand \bibfield  [0]{\@secondoftwo}%
\providecommand \translation [1]{[#1]}%
\providecommand \BibitemOpen [0]{}%
\providecommand \bibitemStop [0]{}%
\providecommand \bibitemNoStop [0]{.\EOS\space}%
\providecommand \EOS [0]{\spacefactor3000\relax}%
\providecommand \BibitemShut  [1]{\csname bibitem#1\endcsname}%
\let\auto@bib@innerbib\@empty
\bibitem [{\citenamefont {Landsteiner}(2016)}]{Landsteiner:1610.04413}%
  \BibitemOpen
  \bibfield  {author} {\bibinfo {author} {\bibfnamefont {K.}~\bibnamefont
  {Landsteiner}},\ }\href {http://dx.doi.org/10.5506/APhysPolB.47.2617}
  {\bibfield  {journal} {\bibinfo  {journal} {Acta~Phys.~Pol.~B}\ }\textbf
  {\bibinfo {volume} {47}},\ \bibinfo {pages} {2617} (\bibinfo {year}
  {2016})},\ \Eprint {https://arxiv.org/abs/1610.04413} {1610.04413}
  \BibitemShut {NoStop}%
\bibitem [{\citenamefont {Kharzeev}\ \emph {et~al.}(2012)\citenamefont
  {Kharzeev}, \citenamefont {Landsteiner}, \citenamefont {Schmitt},\ and\
  \citenamefont {Yee}}]{Kharzeev:1211.6245}%
  \BibitemOpen
  \bibfield  {author} {\bibinfo {author} {\bibfnamefont {D.~E.}\ \bibnamefont
  {Kharzeev}}, \bibinfo {author} {\bibfnamefont {K.}~\bibnamefont
  {Landsteiner}}, \bibinfo {author} {\bibfnamefont {A.}~\bibnamefont
  {Schmitt}},\ and\ \bibinfo {author} {\bibfnamefont {H.}~\bibnamefont {Yee}},\
  }\href {http://dx.doi.org/10.1007/978-3-642-37305-3} {\bibfield  {journal}
  {\bibinfo  {journal} {Lect.~Notes~Phys.}\ }\textbf {\bibinfo {volume}
  {871}},\ \bibinfo {pages} {1} (\bibinfo {year} {2012})},\ \Eprint
  {https://arxiv.org/abs/1211.6245} {1211.6245} \BibitemShut {NoStop}%
\bibitem [{\citenamefont {Adler}(1969)}]{Adler:PhysRev69}%
  \BibitemOpen
  \bibfield  {author} {\bibinfo {author} {\bibfnamefont {S.~L.}\ \bibnamefont
  {Adler}},\ }\href {http://dx.doi.org/10.1103/PhysRev.177.2426} {\bibfield
  {journal} {\bibinfo  {journal} {Phys.~Rev.}\ }\textbf {\bibinfo {volume}
  {177}},\ \bibinfo {pages} {2426 } (\bibinfo {year} {1969})}\BibitemShut
  {NoStop}%
\bibitem [{\citenamefont {Fukushima}\ \emph {et~al.}(2008)\citenamefont
  {Fukushima}, \citenamefont {Kharzeev},\ and\ \citenamefont
  {Warringa}}]{Kharzeev:0808.3382}%
  \BibitemOpen
  \bibfield  {author} {\bibinfo {author} {\bibfnamefont {K.}~\bibnamefont
  {Fukushima}}, \bibinfo {author} {\bibfnamefont {D.~E.}\ \bibnamefont
  {Kharzeev}},\ and\ \bibinfo {author} {\bibfnamefont {H.~J.}\ \bibnamefont
  {Warringa}},\ }\href {https://dx.doi.org/10.1103/PhysRevD.78.074033}
  {\bibfield  {journal} {\bibinfo  {journal} {Phys.~Rev.~D}\ }\textbf {\bibinfo
  {volume} {78}},\ \bibinfo {pages} {074033} (\bibinfo {year} {2008})},\
  \Eprint {https://arxiv.org/abs/0808.3382} {0808.3382} \BibitemShut {NoStop}%
\bibitem [{\citenamefont {Metlitski}\ and\ \citenamefont
  {Zhitnitsky}(2005)}]{Metlitski:hep-ph/0505072}%
  \BibitemOpen
  \bibfield  {author} {\bibinfo {author} {\bibfnamefont {M.~A.}\ \bibnamefont
  {Metlitski}}\ and\ \bibinfo {author} {\bibfnamefont {A.~R.}\ \bibnamefont
  {Zhitnitsky}},\ }\href {http://dx.doi.org/10.1103/PhysRevD.72.045011}
  {\bibfield  {journal} {\bibinfo  {journal} {Phys.~Rev.~D}\ }\textbf {\bibinfo
  {volume} {72}},\ \bibinfo {pages} {045011} (\bibinfo {year} {2005})},\
  \Eprint {https://arxiv.org/abs/hep-ph/0505072} {hep-ph/0505072} \BibitemShut
  {NoStop}%
\bibitem [{\citenamefont {Newman}\ and\ \citenamefont
  {Son}(2006)}]{Son:hep-ph/0510049}%
  \BibitemOpen
  \bibfield  {author} {\bibinfo {author} {\bibfnamefont {G.~M.}\ \bibnamefont
  {Newman}}\ and\ \bibinfo {author} {\bibfnamefont {D.~T.}\ \bibnamefont
  {Son}},\ }\href {http://dx.doi.org/10.1103/PhysRevD.73.045006} {\bibfield
  {journal} {\bibinfo  {journal} {Phys.~Rev.~D}\ }\textbf {\bibinfo {volume}
  {73}},\ \bibinfo {pages} {045006} (\bibinfo {year} {2006})},\ \Eprint
  {https://arxiv.org/abs/hep-ph/0510049} {hep-ph/0510049} \BibitemShut
  {NoStop}%
\bibitem [{\citenamefont {Skokov}\ \emph {et~al.}(2017)\citenamefont {Skokov},
  \citenamefont {Sorensen}, \citenamefont {Koch}, \citenamefont {Schlichting},
  \citenamefont {Thomas}, \citenamefont {Voloshin}, \citenamefont {Wang},\ and\
  \citenamefont {Yee}}]{Skokov:1608.00982}%
  \BibitemOpen
  \bibfield  {author} {\bibinfo {author} {\bibfnamefont {V.}~\bibnamefont
  {Skokov}}, \bibinfo {author} {\bibfnamefont {P.}~\bibnamefont {Sorensen}},
  \bibinfo {author} {\bibfnamefont {V.}~\bibnamefont {Koch}}, \bibinfo {author}
  {\bibfnamefont {S.}~\bibnamefont {Schlichting}}, \bibinfo {author}
  {\bibfnamefont {J.}~\bibnamefont {Thomas}}, \bibinfo {author} {\bibfnamefont
  {S.}~\bibnamefont {Voloshin}}, \bibinfo {author} {\bibfnamefont
  {G.}~\bibnamefont {Wang}},\ and\ \bibinfo {author} {\bibfnamefont
  {H.}~\bibnamefont {Yee}},\ }\href
  {http://dx.doi.org/10.1088/1674-1137/41/7/072001} {\bibfield  {journal}
  {\bibinfo  {journal} {Chin.~Phys.~C}\ }\textbf {\bibinfo {volume} {41}},\
  \bibinfo {pages} {072001} (\bibinfo {year} {2017})},\ \Eprint
  {https://arxiv.org/abs/1608.00982} {1608.00982} \BibitemShut {NoStop}%
\bibitem [{\citenamefont {{CMS Collaboration}}(2018)}]{CMS:1708.01602}%
  \BibitemOpen
  \bibfield  {author} {\bibinfo {author} {\bibnamefont {{CMS Collaboration}}},\
  }\href {http://dx.doi.org/10.1103/PhysRevC.97.044912} {\bibfield  {journal}
  {\bibinfo  {journal} {Phys.~Rev.~C}\ }\textbf {\bibinfo {volume} {97}},\
  \bibinfo {pages} {044912} (\bibinfo {year} {2018})},\ \Eprint
  {https://arxiv.org/abs/1708.01602} {1708.01602} \BibitemShut {NoStop}%
\bibitem [{\citenamefont {Blaschke}\ \emph {et~al.}(2016)\citenamefont
  {Blaschke}, \citenamefont {Aichelin}, \citenamefont {Bratkovskaya},
  \citenamefont {Friese}, \citenamefont {Gazdzicki}, \citenamefont {Randrup},
  \citenamefont {Rogachevsky}, \citenamefont {Teryaev},\ and\ \citenamefont
  {Toneev}}]{NICAWhitePaper}%
  \BibitemOpen
  \bibfield  {author} {\bibinfo {author} {\bibfnamefont {D.}~\bibnamefont
  {Blaschke}}, \bibinfo {author} {\bibfnamefont {J.}~\bibnamefont {Aichelin}},
  \bibinfo {author} {\bibfnamefont {E.}~\bibnamefont {Bratkovskaya}}, \bibinfo
  {author} {\bibfnamefont {V.}~\bibnamefont {Friese}}, \bibinfo {author}
  {\bibfnamefont {M.}~\bibnamefont {Gazdzicki}}, \bibinfo {author}
  {\bibfnamefont {J.}~\bibnamefont {Randrup}}, \bibinfo {author} {\bibfnamefont
  {O.}~\bibnamefont {Rogachevsky}}, \bibinfo {author} {\bibfnamefont
  {O.}~\bibnamefont {Teryaev}},\ and\ \bibinfo {author} {\bibfnamefont
  {V.}~\bibnamefont {Toneev}},\ }\href
  {http://dx.doi.org/10.1140/epja/i2016-16267-x} {\bibfield  {journal}
  {\bibinfo  {journal} {Eur.~Phys.~J.~A}\ }\textbf {\bibinfo {volume} {52}},\
  \bibinfo {pages} {267} (\bibinfo {year} {2016})}\BibitemShut {NoStop}%
\bibitem [{\citenamefont {{J.~Adam et al. (STAR
  Collaboration)}}(2020)}]{Adam:2006.04251}%
  \BibitemOpen
  \bibfield  {author} {\bibinfo {author} {\bibnamefont {{J.~Adam et al. (STAR
  Collaboration)}}},\ }\href@noop {} {\bibinfo {title} {Charge separation
  measurements in {p(d)+Au} and {Au+Au} collisions; implications for the chiral
  magnetic effect}} (\bibinfo {year} {2020}),\ \Eprint
  {https://arxiv.org/abs/2006.04251} {2006.04251} \BibitemShut {NoStop}%
\bibitem [{\citenamefont {Lacey}\ and\ \citenamefont
  {Magdy}(2020)}]{Lacey:2006.04132}%
  \BibitemOpen
  \bibfield  {author} {\bibinfo {author} {\bibfnamefont {R.~A.}\ \bibnamefont
  {Lacey}}\ and\ \bibinfo {author} {\bibfnamefont {N.}~\bibnamefont {Magdy}},\
  }\href@noop {} {\bibinfo {title} {Quantification of the {Chiral Magnetic
  Effect} in {Au+Au} collisions at $\sqrt{s_{NN}} = 200$ gevs}} (\bibinfo
  {year} {2020}),\ \Eprint {https://arxiv.org/abs/2006.04132} {2006.04132}
  \BibitemShut {NoStop}%
\bibitem [{\citenamefont {Zhao}\ and\ \citenamefont
  {Wang}(2019)}]{Zhao:1906.11413}%
  \BibitemOpen
  \bibfield  {author} {\bibinfo {author} {\bibfnamefont {J.}~\bibnamefont
  {Zhao}}\ and\ \bibinfo {author} {\bibfnamefont {F.}~\bibnamefont {Wang}},\
  }\href {http://dx.doi.org/10.1016/j.ppnp.2019.05.001} {\bibfield  {journal}
  {\bibinfo  {journal} {Prog.~Part.~Nucl.~Phys.}\ }\textbf {\bibinfo {volume}
  {107}},\ \bibinfo {pages} {200 } (\bibinfo {year} {2019})},\ \Eprint
  {https://arxiv.org/abs/1906.11413} {1906.11413} \BibitemShut {NoStop}%
\bibitem [{\citenamefont {Huang}\ \emph {et~al.}(2020)\citenamefont {Huang},
  \citenamefont {Nie},\ and\ \citenamefont {Ma}}]{Huang:1906.11631}%
  \BibitemOpen
  \bibfield  {author} {\bibinfo {author} {\bibfnamefont {L.}~\bibnamefont
  {Huang}}, \bibinfo {author} {\bibfnamefont {M.}~\bibnamefont {Nie}},\ and\
  \bibinfo {author} {\bibfnamefont {G.}~\bibnamefont {Ma}},\ }\href
  {http://dx.doi.org/10.1103/PhysRevC.101.024916} {\bibfield  {journal}
  {\bibinfo  {journal} {Phys.~Rev.~C}\ }\textbf {\bibinfo {volume} {101}},\
  \bibinfo {pages} {024916} (\bibinfo {year} {2020})},\ \Eprint
  {https://arxiv.org/abs/1906.11631} {1906.11631} \BibitemShut {NoStop}%
\bibitem [{\citenamefont {Bzdak}\ \emph {et~al.}(2013)\citenamefont {Bzdak},
  \citenamefont {Koch},\ and\ \citenamefont {Liao}}]{Bzdak:1207.7327}%
  \BibitemOpen
  \bibfield  {author} {\bibinfo {author} {\bibfnamefont {A.}~\bibnamefont
  {Bzdak}}, \bibinfo {author} {\bibfnamefont {V.}~\bibnamefont {Koch}},\ and\
  \bibinfo {author} {\bibfnamefont {J.}~\bibnamefont {Liao}},\ }\href
  {http://dx.doi.org/10.1007/978-3-642-37305-3_19} {\bibfield  {journal}
  {\bibinfo  {journal} {Lect.~Notes Phys.}\ }\textbf {\bibinfo {volume}
  {871}},\ \bibinfo {pages} {503 } (\bibinfo {year} {2013})},\ \Eprint
  {https://arxiv.org/abs/1207.7327} {1207.7327} \BibitemShut {NoStop}%
\bibitem [{\citenamefont {Kharzeev}\ and\ \citenamefont
  {Liao}(2019)}]{KharzeevNuclearPhysicsNews}%
  \BibitemOpen
  \bibfield  {author} {\bibinfo {author} {\bibfnamefont {D.~E.}\ \bibnamefont
  {Kharzeev}}\ and\ \bibinfo {author} {\bibfnamefont {J.}~\bibnamefont
  {Liao}},\ }\href {http://dx.doi.org/10.1080/10619127.2018.1495479} {\bibfield
   {journal} {\bibinfo  {journal} {Nucl.~Phys.~News}\ }\textbf {\bibinfo
  {volume} {29}},\ \bibinfo {pages} {26 } (\bibinfo {year} {2019})}\BibitemShut
  {NoStop}%
\bibitem [{\citenamefont {{STAR Collaboration: J. Adam et
  al.}}(2019)}]{STAR:1911.00596}%
  \BibitemOpen
  \bibfield  {author} {\bibinfo {author} {\bibnamefont {{STAR Collaboration: J.
  Adam et al.}}},\ }\href@noop {} {\bibinfo {title} {Methods for a blind
  analysis of isobar data collected by the {STAR} collaboration}} (\bibinfo
  {year} {2019}),\ \Eprint {https://arxiv.org/abs/1911.00596} {1911.00596}
  \BibitemShut {NoStop}%
\bibitem [{\citenamefont {Teaney}(2010)}]{Teaney:0905.2433}%
  \BibitemOpen
  \bibfield  {author} {\bibinfo {author} {\bibfnamefont {D.~A.}\ \bibnamefont
  {Teaney}},\ }\href {http://dx.doi.org/10.1142/9789814293297_0004} {\bibfield
  {journal} {\bibinfo  {journal} {Quark-Gluon Plasma}\ }\textbf {\bibinfo
  {volume} {4}},\ \bibinfo {pages} {207 } (\bibinfo {year} {2010})},\ \Eprint
  {https://arxiv.org/abs/0905.2433} {0905.2433} \BibitemShut {NoStop}%
\bibitem [{\citenamefont {{S.~A.~Voloshin~(STAR
  Collaboration)}}(2011)}]{Voloshin:0806.0029}%
  \BibitemOpen
  \bibfield  {author} {\bibinfo {author} {\bibnamefont {{S.~A.~Voloshin~(STAR
  Collaboration)}}},\ }\href {http://dx.doi.org/10.1007/s12648-011-0137-0}
  {\bibfield  {journal} {\bibinfo  {journal} {Indian~J.~Phys.}\ }\textbf
  {\bibinfo {volume} {85}},\ \bibinfo {pages} {1103 } (\bibinfo {year}
  {2011})},\ \Eprint {https://arxiv.org/abs/0806.0029} {0806.0029} \BibitemShut
  {NoStop}%
\bibitem [{\citenamefont {Becattini}\ \emph {et~al.}(2017)\citenamefont
  {Becattini}, \citenamefont {Karpenko}, \citenamefont {Lisa}, \citenamefont
  {Upsal},\ and\ \citenamefont {Voloshin}}]{Upsal:1610.02506}%
  \BibitemOpen
  \bibfield  {author} {\bibinfo {author} {\bibfnamefont {F.}~\bibnamefont
  {Becattini}}, \bibinfo {author} {\bibfnamefont {I.}~\bibnamefont {Karpenko}},
  \bibinfo {author} {\bibfnamefont {M.}~\bibnamefont {Lisa}}, \bibinfo {author}
  {\bibfnamefont {I.}~\bibnamefont {Upsal}},\ and\ \bibinfo {author}
  {\bibfnamefont {S.}~\bibnamefont {Voloshin}},\ }\href
  {http://dx.doi.org/10.1103/PhysRevC.95.054902} {\bibfield  {journal}
  {\bibinfo  {journal} {Phys.~Rev.~C}\ }\textbf {\bibinfo {volume} {95}},\
  \bibinfo {pages} {054902} (\bibinfo {year} {2017})},\ \Eprint
  {https://arxiv.org/abs/1610.02506} {1610.02506} \BibitemShut {NoStop}%
\bibitem [{\citenamefont {Jiang}\ \emph {et~al.}(2018)\citenamefont {Jiang},
  \citenamefont {Shi}, \citenamefont {Yin},\ and\ \citenamefont
  {Liao}}]{Liao:1611.04586}%
  \BibitemOpen
  \bibfield  {author} {\bibinfo {author} {\bibfnamefont {Y.}~\bibnamefont
  {Jiang}}, \bibinfo {author} {\bibfnamefont {S.}~\bibnamefont {Shi}}, \bibinfo
  {author} {\bibfnamefont {Y.}~\bibnamefont {Yin}},\ and\ \bibinfo {author}
  {\bibfnamefont {J.}~\bibnamefont {Liao}},\ }\href
  {http://dx.doi.org/10.1088/1674-1137/42/1/011001} {\bibfield  {journal}
  {\bibinfo  {journal} {Chin.~Phys.~C}\ }\textbf {\bibinfo {volume} {42}},\
  \bibinfo {pages} {011001} (\bibinfo {year} {2018})},\ \Eprint
  {https://arxiv.org/abs/1611.04586} {1611.04586} \BibitemShut {NoStop}%
\bibitem [{\citenamefont {Shi}\ \emph {et~al.}(2018{\natexlab{a}})\citenamefont
  {Shi}, \citenamefont {Zhang}, \citenamefont {Hou},\ and\ \citenamefont
  {Liao}}]{Liao:1807.05604}%
  \BibitemOpen
  \bibfield  {author} {\bibinfo {author} {\bibfnamefont {S.}~\bibnamefont
  {Shi}}, \bibinfo {author} {\bibfnamefont {H.}~\bibnamefont {Zhang}}, \bibinfo
  {author} {\bibfnamefont {D.}~\bibnamefont {Hou}},\ and\ \bibinfo {author}
  {\bibfnamefont {J.}~\bibnamefont {Liao}},\ }\href
  {http://dx.doi.org/10.1016/j.nuclphysa.2018.10.007} {\bibfield  {journal}
  {\bibinfo  {journal} {Nucl.~Phys.~A}\ }\textbf {\bibinfo {volume} {982}},\
  \bibinfo {pages} {539 } (\bibinfo {year} {2018}{\natexlab{a}})},\ \Eprint
  {https://arxiv.org/abs/1807.05604} {1807.05604} \BibitemShut {NoStop}%
\bibitem [{\citenamefont {Shi}\ \emph {et~al.}(2018{\natexlab{b}})\citenamefont
  {Shi}, \citenamefont {Jiang}, \citenamefont {Lilleskov},\ and\ \citenamefont
  {Liao}}]{Lilleskov:1711.02496}%
  \BibitemOpen
  \bibfield  {author} {\bibinfo {author} {\bibfnamefont {S.}~\bibnamefont
  {Shi}}, \bibinfo {author} {\bibfnamefont {Y.}~\bibnamefont {Jiang}}, \bibinfo
  {author} {\bibfnamefont {E.}~\bibnamefont {Lilleskov}},\ and\ \bibinfo
  {author} {\bibfnamefont {J.}~\bibnamefont {Liao}},\ }\href
  {http://dx.doi.org/10.1016/j.aop.2018.04.026} {\bibfield  {journal} {\bibinfo
   {journal} {Ann.~Phys.}\ }\textbf {\bibinfo {volume} {394}},\ \bibinfo
  {pages} {50 } (\bibinfo {year} {2018}{\natexlab{b}})},\ \Eprint
  {https://arxiv.org/abs/1711.02496} {1711.02496} \BibitemShut {NoStop}%
\bibitem [{\citenamefont {Erdmenger}\ \emph {et~al.}(2009)\citenamefont
  {Erdmenger}, \citenamefont {Haack}, \citenamefont {Kaminski},\ and\
  \citenamefont {Yarom}}]{Erdmenger:0809.2488}%
  \BibitemOpen
  \bibfield  {author} {\bibinfo {author} {\bibfnamefont {J.}~\bibnamefont
  {Erdmenger}}, \bibinfo {author} {\bibfnamefont {M.}~\bibnamefont {Haack}},
  \bibinfo {author} {\bibfnamefont {M.}~\bibnamefont {Kaminski}},\ and\
  \bibinfo {author} {\bibfnamefont {A.}~\bibnamefont {Yarom}},\ }\href
  {http://dx.doi.org/10.1088/1126-6708/2009/01/055} {\bibfield  {journal}
  {\bibinfo  {journal} {JHEP}\ }\textbf {\bibinfo {volume} {0901}},\ \bibinfo
  {pages} {055}},\ \Eprint {https://arxiv.org/abs/0809.2488} {0809.2488}
  \BibitemShut {NoStop}%
\bibitem [{\citenamefont {Son}\ and\ \citenamefont
  {Surowka}(2009)}]{Son:0906.5044}%
  \BibitemOpen
  \bibfield  {author} {\bibinfo {author} {\bibfnamefont {D.~T.}\ \bibnamefont
  {Son}}\ and\ \bibinfo {author} {\bibfnamefont {P.}~\bibnamefont {Surowka}},\
  }\href {10.1103/PhysRevLett.103.191601} {\bibfield  {journal} {\bibinfo
  {journal} {Phys.~Rev.~Lett.}\ }\textbf {\bibinfo {volume} {103}},\ \bibinfo
  {pages} {191601} (\bibinfo {year} {2009})},\ \Eprint
  {https://arxiv.org/abs/0906.5044} {0906.5044} \BibitemShut {NoStop}%
\bibitem [{\citenamefont {Banerjee}\ \emph {et~al.}(2011)\citenamefont
  {Banerjee}, \citenamefont {Bhattacharya}, \citenamefont {Bhattacharyya},
  \citenamefont {Dutta}, \citenamefont {Loganayagam},\ and\ \citenamefont
  {Sur\'{o}wka}}]{Surowka:0809.2596}%
  \BibitemOpen
  \bibfield  {author} {\bibinfo {author} {\bibfnamefont {N.}~\bibnamefont
  {Banerjee}}, \bibinfo {author} {\bibfnamefont {J.}~\bibnamefont
  {Bhattacharya}}, \bibinfo {author} {\bibfnamefont {S.}~\bibnamefont
  {Bhattacharyya}}, \bibinfo {author} {\bibfnamefont {S.}~\bibnamefont
  {Dutta}}, \bibinfo {author} {\bibfnamefont {R.}~\bibnamefont {Loganayagam}},\
  and\ \bibinfo {author} {\bibfnamefont {P.}~\bibnamefont {Sur\'{o}wka}},\
  }\href@noop {} {\bibfield  {journal} {\bibinfo  {journal} {JHEP}\ }\textbf
  {\bibinfo {volume} {1101}},\ \bibinfo {pages} {094}},\ \Eprint
  {https://arxiv.org/abs/0809.2596} {0809.2596} \BibitemShut {NoStop}%
\bibitem [{\citenamefont {Guo}\ \emph {et~al.}(2017)\citenamefont {Guo},
  \citenamefont {Kharzeev}, \citenamefont {Huang}, \citenamefont {Deng},\ and\
  \citenamefont {Hirono}}]{Hirono:1704.05375}%
  \BibitemOpen
  \bibfield  {author} {\bibinfo {author} {\bibfnamefont {X.}~\bibnamefont
  {Guo}}, \bibinfo {author} {\bibfnamefont {D.~E.}\ \bibnamefont {Kharzeev}},
  \bibinfo {author} {\bibfnamefont {X.}~\bibnamefont {Huang}}, \bibinfo
  {author} {\bibfnamefont {W.}~\bibnamefont {Deng}},\ and\ \bibinfo {author}
  {\bibfnamefont {Y.}~\bibnamefont {Hirono}},\ }\href
  {http://dx.doi.org/10.1016/j.nuclphysa.2017.06.039} {\bibfield  {journal}
  {\bibinfo  {journal} {Nucl.~Phys.~A}\ }\textbf {\bibinfo {volume} {967}},\
  \bibinfo {pages} {1} (\bibinfo {year} {2017})},\ \Eprint
  {https://arxiv.org/abs/1704.05375} {1704.05375} \BibitemShut {NoStop}%
\bibitem [{\citenamefont {Inghirami}\ \emph {et~al.}(2020)\citenamefont
  {Inghirami}, \citenamefont {Mace}, \citenamefont {Hirono}, \citenamefont
  {{Del Zanna}}, \citenamefont {Kharzeev},\ and\ \citenamefont
  {Bleicher}}]{Kharzeev:1908.07605}%
  \BibitemOpen
  \bibfield  {author} {\bibinfo {author} {\bibfnamefont {G.}~\bibnamefont
  {Inghirami}}, \bibinfo {author} {\bibfnamefont {M.}~\bibnamefont {Mace}},
  \bibinfo {author} {\bibfnamefont {Y.}~\bibnamefont {Hirono}}, \bibinfo
  {author} {\bibfnamefont {L.}~\bibnamefont {{Del Zanna}}}, \bibinfo {author}
  {\bibfnamefont {D.~E.}\ \bibnamefont {Kharzeev}},\ and\ \bibinfo {author}
  {\bibfnamefont {M.}~\bibnamefont {Bleicher}},\ }\href {http://dx.doi.org/}
  {\bibfield  {journal} {\bibinfo  {journal} {Eur.~Phys.~J.~C}\ }\textbf
  {\bibinfo {volume} {80}},\ \bibinfo {pages} {293} (\bibinfo {year} {2020})},\
  \Eprint {https://arxiv.org/abs/1908.07605} {1908.07605} \BibitemShut
  {NoStop}%
\bibitem [{\citenamefont {Liang}\ \emph {et~al.}(2020)\citenamefont {Liang},
  \citenamefont {Liao}, \citenamefont {Lin}, \citenamefont {Yan},\ and\
  \citenamefont {Li}}]{Liao:2004.04440}%
  \BibitemOpen
  \bibfield  {author} {\bibinfo {author} {\bibfnamefont {G.}~\bibnamefont
  {Liang}}, \bibinfo {author} {\bibfnamefont {J.}~\bibnamefont {Liao}},
  \bibinfo {author} {\bibfnamefont {S.}~\bibnamefont {Lin}}, \bibinfo {author}
  {\bibfnamefont {L.}~\bibnamefont {Yan}},\ and\ \bibinfo {author}
  {\bibfnamefont {M.}~\bibnamefont {Li}},\ }\href@noop {} {\bibinfo {title}
  {Chiral magnetic effect in isobar collisions from stochastic hydrodynamics}}
  (\bibinfo {year} {2020}),\ \Eprint {https://arxiv.org/abs/2004.04440}
  {2004.04440} \BibitemShut {NoStop}%
\bibitem [{\citenamefont {Sadofyev}\ and\ \citenamefont
  {Isachenkov}(2011)}]{Sadofyev:1010.1550}%
  \BibitemOpen
  \bibfield  {author} {\bibinfo {author} {\bibfnamefont {A.~V.}\ \bibnamefont
  {Sadofyev}}\ and\ \bibinfo {author} {\bibfnamefont {M.~V.}\ \bibnamefont
  {Isachenkov}},\ }\href {http://dx.doi.org/10.1016/j.physletb.2011.02.041}
  {\bibfield  {journal} {\bibinfo  {journal} {Phys.~Lett.~B}\ }\textbf
  {\bibinfo {volume} {697}},\ \bibinfo {pages} {404 } (\bibinfo {year}
  {2011})},\ \Eprint {https://arxiv.org/abs/1010.1550} {1010.1550} \BibitemShut
  {NoStop}%
\bibitem [{\citenamefont {Gorbar}\ \emph {et~al.}(2013)\citenamefont {Gorbar},
  \citenamefont {Miransky}, \citenamefont {Shovkovy},\ and\ \citenamefont
  {Wang}}]{Miransky:1304.4606}%
  \BibitemOpen
  \bibfield  {author} {\bibinfo {author} {\bibfnamefont {E.~V.}\ \bibnamefont
  {Gorbar}}, \bibinfo {author} {\bibfnamefont {V.~A.}\ \bibnamefont
  {Miransky}}, \bibinfo {author} {\bibfnamefont {I.~A.}\ \bibnamefont
  {Shovkovy}},\ and\ \bibinfo {author} {\bibfnamefont {X.}~\bibnamefont
  {Wang}},\ }\href {http://dx.doi.org/10.1103/PhysRevD.88.025025} {\bibfield
  {journal} {\bibinfo  {journal} {Phys.~Rev.~D}\ }\textbf {\bibinfo {volume}
  {88}},\ \bibinfo {pages} {025025} (\bibinfo {year} {2013})},\ \Eprint
  {https://arxiv.org/abs/1304.4606} {1304.4606} \BibitemShut {NoStop}%
\bibitem [{\citenamefont {Golkar}\ and\ \citenamefont
  {Son}(2012)}]{Son:1207.5806}%
  \BibitemOpen
  \bibfield  {author} {\bibinfo {author} {\bibfnamefont {S.}~\bibnamefont
  {Golkar}}\ and\ \bibinfo {author} {\bibfnamefont {D.~T.}\ \bibnamefont
  {Son}},\ }\href {http://dx.doi.org/10.1007/JHEP02(2015)169} {\bibfield
  {journal} {\bibinfo  {journal} {JHEP}\ }\textbf {\bibinfo {volume} {02}},\
  \bibinfo {pages} {169}},\ \Eprint {https://arxiv.org/abs/1207.5806}
  {1207.5806} \BibitemShut {NoStop}%
\bibitem [{\citenamefont {Jensen}\ \emph {et~al.}(2013)\citenamefont {Jensen},
  \citenamefont {Kovtun},\ and\ \citenamefont {Ritz}}]{Jensen:1307.3234}%
  \BibitemOpen
  \bibfield  {author} {\bibinfo {author} {\bibfnamefont {K.}~\bibnamefont
  {Jensen}}, \bibinfo {author} {\bibfnamefont {P.}~\bibnamefont {Kovtun}},\
  and\ \bibinfo {author} {\bibfnamefont {A.}~\bibnamefont {Ritz}},\ }\href
  {http://dx.doi.org/10.1007/JHEP10(2013)186} {\bibfield  {journal} {\bibinfo
  {journal} {JHEP}\ }\textbf {\bibinfo {volume} {1310}},\ \bibinfo {pages}
  {186}},\ \Eprint {https://arxiv.org/abs/1307.3234} {1307.3234} \BibitemShut
  {NoStop}%
\bibitem [{\citenamefont {Gursoy}\ and\ \citenamefont
  {Jansen}(2014)}]{Gursoy:1407.3282}%
  \BibitemOpen
  \bibfield  {author} {\bibinfo {author} {\bibfnamefont {U.}~\bibnamefont
  {Gursoy}}\ and\ \bibinfo {author} {\bibfnamefont {A.}~\bibnamefont
  {Jansen}},\ }\href {http://dx.doi.org/10.1007/JHEP10(2014)092} {\bibfield
  {journal} {\bibinfo  {journal} {JHEP}\ }\textbf {\bibinfo {volume} {1410}},\
  \bibinfo {pages} {92}},\ \Eprint {https://arxiv.org/abs/1407.3282}
  {1407.3282} \BibitemShut {NoStop}%
\bibitem [{\citenamefont {{Jimenez-Alba}}\ and\ \citenamefont
  {Melgar}(2014)}]{Melgar:1404.2434}%
  \BibitemOpen
  \bibfield  {author} {\bibinfo {author} {\bibfnamefont {A.}~\bibnamefont
  {{Jimenez-Alba}}}\ and\ \bibinfo {author} {\bibfnamefont {L.}~\bibnamefont
  {Melgar}},\ }\href {http://dx.doi.org/10.1007/JHEP10(2014)120} {\bibfield
  {journal} {\bibinfo  {journal} {JHEP}\ }\textbf {\bibinfo {volume} {10}},\
  \bibinfo {pages} {120}},\ \Eprint {https://arxiv.org/abs/1404.2434}
  {1404.2434} \BibitemShut {NoStop}%
\bibitem [{\citenamefont {Avdoshkin}\ \emph {et~al.}(2018)\citenamefont
  {Avdoshkin}, \citenamefont {Sadofyev},\ and\ \citenamefont
  {Zakharov}}]{Avdoshkin:1712.01256}%
  \BibitemOpen
  \bibfield  {author} {\bibinfo {author} {\bibfnamefont {A.}~\bibnamefont
  {Avdoshkin}}, \bibinfo {author} {\bibfnamefont {A.~V.}\ \bibnamefont
  {Sadofyev}},\ and\ \bibinfo {author} {\bibfnamefont {V.~I.}\ \bibnamefont
  {Zakharov}},\ }\href {http://dx.doi.org/10.1103/PhysRevD.97.085020}
  {\bibfield  {journal} {\bibinfo  {journal} {Phys.~Rev.~D}\ }\textbf {\bibinfo
  {volume} {97}},\ \bibinfo {pages} {085020} (\bibinfo {year} {2018})},\
  \Eprint {https://arxiv.org/abs/1712.01256} {1712.01256} \BibitemShut
  {NoStop}%
\bibitem [{\citenamefont {Yamamoto}(2011)}]{Yamamoto:1105.0385}%
  \BibitemOpen
  \bibfield  {author} {\bibinfo {author} {\bibfnamefont {A.}~\bibnamefont
  {Yamamoto}},\ }\href {http://dx.doi.org/10.1103/PhysRevLett.107.031601}
  {\bibfield  {journal} {\bibinfo  {journal} {Phys.~Rev.~Lett.}\ }\textbf
  {\bibinfo {volume} {107}},\ \bibinfo {pages} {031601} (\bibinfo {year}
  {2011})},\ \Eprint {https://arxiv.org/abs/1105.0385} {1105.0385} \BibitemShut
  {NoStop}%
\bibitem [{\citenamefont {Braguta}\ \emph {et~al.}(2014)\citenamefont
  {Braguta}, \citenamefont {Chernodub}, \citenamefont {Goy}, \citenamefont
  {Landsteiner}, \citenamefont {Molochkov},\ and\ \citenamefont
  {Polikarpov}}]{Braguta:1401.8095}%
  \BibitemOpen
  \bibfield  {author} {\bibinfo {author} {\bibfnamefont {V.}~\bibnamefont
  {Braguta}}, \bibinfo {author} {\bibfnamefont {M.~N.}\ \bibnamefont
  {Chernodub}}, \bibinfo {author} {\bibfnamefont {V.~A.}\ \bibnamefont {Goy}},
  \bibinfo {author} {\bibfnamefont {K.}~\bibnamefont {Landsteiner}}, \bibinfo
  {author} {\bibfnamefont {A.~V.}\ \bibnamefont {Molochkov}},\ and\ \bibinfo
  {author} {\bibfnamefont {M.~I.}\ \bibnamefont {Polikarpov}},\ }\href
  {http://dx.doi.org/10.1103/PhysRevD.89.074510} {\bibfield  {journal}
  {\bibinfo  {journal} {Phys.~Rev.~D}\ }\textbf {\bibinfo {volume} {89}},\
  \bibinfo {pages} {074510} (\bibinfo {year} {2014})},\ \Eprint
  {https://arxiv.org/abs/1401.8095} {1401.8095} \BibitemShut {NoStop}%
\bibitem [{\citenamefont {Puhr}\ and\ \citenamefont
  {Buividovich}(2017)}]{Buividovich:16:6}%
  \BibitemOpen
  \bibfield  {author} {\bibinfo {author} {\bibfnamefont {M.}~\bibnamefont
  {Puhr}}\ and\ \bibinfo {author} {\bibfnamefont {P.~V.}\ \bibnamefont
  {Buividovich}},\ }\href@noop {} {\bibfield  {journal} {\bibinfo  {journal}
  {Phys.~Rev.~Lett.}\ }\textbf {\bibinfo {volume} {118}},\ \bibinfo {pages}
  {192003} (\bibinfo {year} {2017})},\ \Eprint
  {https://arxiv.org/abs/1611.07263} {1611.07263} \BibitemShut {NoStop}%
\bibitem [{\citenamefont {Burnier}\ \emph {et~al.}(2011)\citenamefont
  {Burnier}, \citenamefont {Kharzeev}, \citenamefont {Liao},\ and\
  \citenamefont {Yee}}]{Kharzeev:1103.1307}%
  \BibitemOpen
  \bibfield  {author} {\bibinfo {author} {\bibfnamefont {Y.}~\bibnamefont
  {Burnier}}, \bibinfo {author} {\bibfnamefont {D.~E.}\ \bibnamefont
  {Kharzeev}}, \bibinfo {author} {\bibfnamefont {J.}~\bibnamefont {Liao}},\
  and\ \bibinfo {author} {\bibfnamefont {H.}~\bibnamefont {Yee}},\ }\href
  {https://dx.doi.org/10.1103/PhysRevLett.107.052303} {\bibfield  {journal}
  {\bibinfo  {journal} {Phys.~Rev.~Lett.}\ }\textbf {\bibinfo {volume} {107}},\
  \bibinfo {pages} {052303} (\bibinfo {year} {2011})},\ \Eprint
  {https://arxiv.org/abs/1103.1307} {1103.1307} \BibitemShut {NoStop}%
\bibitem [{\citenamefont {Kharzeev}\ and\ \citenamefont
  {Yee}(2011)}]{Kharzeev:1012.6026}%
  \BibitemOpen
  \bibfield  {author} {\bibinfo {author} {\bibfnamefont {D.~E.}\ \bibnamefont
  {Kharzeev}}\ and\ \bibinfo {author} {\bibfnamefont {H.}~\bibnamefont {Yee}},\
  }\href {https://dx.doi.org/10.1103/PhysRevD.83.085007} {\bibfield  {journal}
  {\bibinfo  {journal} {Phys.~Rev.~D}\ }\textbf {\bibinfo {volume} {83}},\
  \bibinfo {pages} {085007} (\bibinfo {year} {2011})},\ \Eprint
  {https://arxiv.org/abs/1012.6026} {1012.6026} \BibitemShut {NoStop}%
\bibitem [{\citenamefont {Gattringer}\ and\ \citenamefont
  {Langfeld}(2016)}]{Gattringer:1603.09517}%
  \BibitemOpen
  \bibfield  {author} {\bibinfo {author} {\bibfnamefont {C.}~\bibnamefont
  {Gattringer}}\ and\ \bibinfo {author} {\bibfnamefont {K.}~\bibnamefont
  {Langfeld}},\ }\href {https://dx.doi.org/10.1142/S0217751X16430077}
  {\bibfield  {journal} {\bibinfo  {journal} {Int.~J.~Mod.~Phys.~A}\ }\textbf
  {\bibinfo {volume} {31}},\ \bibinfo {pages} {1643007} (\bibinfo {year}
  {2016})},\ \Eprint {https://arxiv.org/abs/1603.09517} {1603.09517}
  \BibitemShut {NoStop}%
\bibitem [{\citenamefont {Kogut}\ \emph {et~al.}(2001)\citenamefont {Kogut},
  \citenamefont {Sinclair}, \citenamefont {Hands},\ and\ \citenamefont
  {Morrison}}]{Kogut:hep-lat/0105026}%
  \BibitemOpen
  \bibfield  {author} {\bibinfo {author} {\bibfnamefont {J.~B.}\ \bibnamefont
  {Kogut}}, \bibinfo {author} {\bibfnamefont {D.~K.}\ \bibnamefont {Sinclair}},
  \bibinfo {author} {\bibfnamefont {S.~J.}\ \bibnamefont {Hands}},\ and\
  \bibinfo {author} {\bibfnamefont {S.~E.}\ \bibnamefont {Morrison}},\ }\href
  {http://dx.doi.org/10.1103/PhysRevD.64.094505} {\bibfield  {journal}
  {\bibinfo  {journal} {Phys.~Rev.~D}\ }\textbf {\bibinfo {volume} {64}},\
  \bibinfo {pages} {094505} (\bibinfo {year} {2001})},\ \Eprint
  {https://arxiv.org/abs/hep-lat/0105026} {hep-lat/0105026} \BibitemShut
  {NoStop}%
\bibitem [{\citenamefont {Kogut}\ \emph {et~al.}(2000)\citenamefont {Kogut},
  \citenamefont {Stephanov}, \citenamefont {Toublan}, \citenamefont
  {Verbaarschot},\ and\ \citenamefont {Zhitnitsky}}]{Kogut:hep-ph/0001171}%
  \BibitemOpen
  \bibfield  {author} {\bibinfo {author} {\bibfnamefont {J.~B.}\ \bibnamefont
  {Kogut}}, \bibinfo {author} {\bibfnamefont {M.~A.}\ \bibnamefont
  {Stephanov}}, \bibinfo {author} {\bibfnamefont {D.}~\bibnamefont {Toublan}},
  \bibinfo {author} {\bibfnamefont {J.~J.~M.}\ \bibnamefont {Verbaarschot}},\
  and\ \bibinfo {author} {\bibfnamefont {A.}~\bibnamefont {Zhitnitsky}},\
  }\href {http://dx.doi.org/10.1016/S0550-3213(00)00242-X} {\bibfield
  {journal} {\bibinfo  {journal} {Nucl.~Phys.~B}\ }\textbf {\bibinfo {volume}
  {582}},\ \bibinfo {pages} {477 } (\bibinfo {year} {2000})},\ \Eprint
  {https://arxiv.org/abs/hep-ph/0001171} {hep-ph/0001171} \BibitemShut
  {NoStop}%
\bibitem [{\citenamefont {Buividovich}\ \emph {et~al.}(2020)\citenamefont
  {Buividovich}, \citenamefont {{von~Smekal}},\ and\ \citenamefont
  {Smith}}]{Buividovich:20:1}%
  \BibitemOpen
  \bibfield  {author} {\bibinfo {author} {\bibfnamefont {P.~V.}\ \bibnamefont
  {Buividovich}}, \bibinfo {author} {\bibfnamefont {L.}~\bibnamefont
  {{von~Smekal}}},\ and\ \bibinfo {author} {\bibfnamefont {D.}~\bibnamefont
  {Smith}},\ }\href {http://dx.doi.org/10.1103/PhysRevD.102.094510} {\bibfield
  {journal} {\bibinfo  {journal} {Phys.~Rev.~D}\ }\textbf {\bibinfo {volume}
  {102}},\ \bibinfo {pages} {094510} (\bibinfo {year} {2020})},\ \Eprint
  {https://arxiv.org/abs/2007.05639} {2007.05639} \BibitemShut {NoStop}%
\bibitem [{\citenamefont {Boz}\ \emph {et~al.}(2020)\citenamefont {Boz},
  \citenamefont {Giudice}, \citenamefont {Hands},\ and\ \citenamefont
  {Skullerud}}]{Hands:1912.10975}%
  \BibitemOpen
  \bibfield  {author} {\bibinfo {author} {\bibfnamefont {T.}~\bibnamefont
  {Boz}}, \bibinfo {author} {\bibfnamefont {P.}~\bibnamefont {Giudice}},
  \bibinfo {author} {\bibfnamefont {S.}~\bibnamefont {Hands}},\ and\ \bibinfo
  {author} {\bibfnamefont {J.}~\bibnamefont {Skullerud}},\ }\href
  {http://dx.doi.org/10.1103/PhysRevD.101.074506} {\bibfield  {journal}
  {\bibinfo  {journal} {Phys.~Rev.~D}\ }\textbf {\bibinfo {volume} {101}},\
  \bibinfo {pages} {074506} (\bibinfo {year} {2020})},\ \Eprint
  {https://arxiv.org/abs/1912.10975} {1912.10975} \BibitemShut {NoStop}%
\bibitem [{\citenamefont {Wilhelm}\ \emph {et~al.}(2019)\citenamefont
  {Wilhelm}, \citenamefont {Holicki}, \citenamefont {Smith}, \citenamefont
  {Wellegehausen},\ and\ \citenamefont {{von Smekal}}}]{Smith:1910.04495}%
  \BibitemOpen
  \bibfield  {author} {\bibinfo {author} {\bibfnamefont {J.}~\bibnamefont
  {Wilhelm}}, \bibinfo {author} {\bibfnamefont {L.}~\bibnamefont {Holicki}},
  \bibinfo {author} {\bibfnamefont {D.}~\bibnamefont {Smith}}, \bibinfo
  {author} {\bibfnamefont {B.}~\bibnamefont {Wellegehausen}},\ and\ \bibinfo
  {author} {\bibfnamefont {L.}~\bibnamefont {{von Smekal}}},\ }\href
  {http://dx.doi.org/10.1103/PhysRevD.100.114507} {\bibfield  {journal}
  {\bibinfo  {journal} {Phys.~Rev.~D}\ }\textbf {\bibinfo {volume} {100}},\
  \bibinfo {pages} {114507} (\bibinfo {year} {2019})},\ \Eprint
  {https://arxiv.org/abs/1910.04495} {1910.04495} \BibitemShut {NoStop}%
\bibitem [{\citenamefont {Holicki}\ \emph {et~al.}(2017)\citenamefont
  {Holicki}, \citenamefont {Wilhelm}, \citenamefont {Smith}, \citenamefont
  {Wellegehausen},\ and\ \citenamefont {{von Smekal}}}]{Holicki:1701.04664}%
  \BibitemOpen
  \bibfield  {author} {\bibinfo {author} {\bibfnamefont {L.}~\bibnamefont
  {Holicki}}, \bibinfo {author} {\bibfnamefont {J.}~\bibnamefont {Wilhelm}},
  \bibinfo {author} {\bibfnamefont {D.}~\bibnamefont {Smith}}, \bibinfo
  {author} {\bibfnamefont {B.}~\bibnamefont {Wellegehausen}},\ and\ \bibinfo
  {author} {\bibfnamefont {L.}~\bibnamefont {{von Smekal}}},\ }\href
  {https://pos.sissa.it/256/052/} {\bibfield  {journal} {\bibinfo  {journal}
  {PoS}\ }\textbf {\bibinfo {volume} {LATTICE2016}},\ \bibinfo {pages} {052}
  (\bibinfo {year} {2017})},\ \Eprint {https://arxiv.org/abs/1701.04664}
  {1701.04664} \BibitemShut {NoStop}%
\bibitem [{\citenamefont {Boz}\ \emph {et~al.}(2016)\citenamefont {Boz},
  \citenamefont {Giudice}, \citenamefont {Hands}, \citenamefont {Skullerud},\
  and\ \citenamefont {Williams}}]{Hands:1502.01219}%
  \BibitemOpen
  \bibfield  {author} {\bibinfo {author} {\bibfnamefont {T.}~\bibnamefont
  {Boz}}, \bibinfo {author} {\bibfnamefont {P.}~\bibnamefont {Giudice}},
  \bibinfo {author} {\bibfnamefont {S.}~\bibnamefont {Hands}}, \bibinfo
  {author} {\bibfnamefont {J.}~\bibnamefont {Skullerud}},\ and\ \bibinfo
  {author} {\bibfnamefont {A.~G.}\ \bibnamefont {Williams}},\ }\href
  {http://dx.doi.org/10.1063/1.4938682} {\bibfield  {journal} {\bibinfo
  {journal} {AIP~Conf.~Proc.}\ }\textbf {\bibinfo {volume} {1701}},\ \bibinfo
  {pages} {060019} (\bibinfo {year} {2016})},\ \Eprint
  {https://arxiv.org/abs/1502.01219} {1502.01219} \BibitemShut {NoStop}%
\bibitem [{\citenamefont {Cotter}\ \emph {et~al.}(2013)\citenamefont {Cotter},
  \citenamefont {Giudice}, \citenamefont {Hands},\ and\ \citenamefont
  {Skullerud}}]{Hands:1210.4496}%
  \BibitemOpen
  \bibfield  {author} {\bibinfo {author} {\bibfnamefont {S.}~\bibnamefont
  {Cotter}}, \bibinfo {author} {\bibfnamefont {P.}~\bibnamefont {Giudice}},
  \bibinfo {author} {\bibfnamefont {S.}~\bibnamefont {Hands}},\ and\ \bibinfo
  {author} {\bibfnamefont {J.}~\bibnamefont {Skullerud}},\ }\href
  {http://dx.doi.org/10.1103/PhysRevD.87.034507} {\bibfield  {journal}
  {\bibinfo  {journal} {Phys.~Rev.~D}\ }\textbf {\bibinfo {volume} {87}},\
  \bibinfo {pages} {034507} (\bibinfo {year} {2013})},\ \Eprint
  {https://arxiv.org/abs/1210.4496} {1210.4496} \BibitemShut {NoStop}%
\bibitem [{\citenamefont {Strodthoff}\ \emph {et~al.}(2012)\citenamefont
  {Strodthoff}, \citenamefont {Schaefer},\ and\ \citenamefont {{von
  Smekal}}}]{Smekal:1112.5401}%
  \BibitemOpen
  \bibfield  {author} {\bibinfo {author} {\bibfnamefont {N.}~\bibnamefont
  {Strodthoff}}, \bibinfo {author} {\bibfnamefont {B.}~\bibnamefont
  {Schaefer}},\ and\ \bibinfo {author} {\bibfnamefont {L.}~\bibnamefont {{von
  Smekal}}},\ }\href {http://dx.doi.org/10.1103/PhysRevD.85.074007} {\bibfield
  {journal} {\bibinfo  {journal} {Phys.~Rev.~D}\ }\textbf {\bibinfo {volume}
  {85}},\ \bibinfo {pages} {074007} (\bibinfo {year} {2012})},\ \Eprint
  {https://arxiv.org/abs/1112.5401} {1112.5401} \BibitemShut {NoStop}%
\bibitem [{\citenamefont {Strodthoff}\ and\ \citenamefont {{von
  Smekal}}(2014)}]{Strodthoff:1306.2897}%
  \BibitemOpen
  \bibfield  {author} {\bibinfo {author} {\bibfnamefont {N.}~\bibnamefont
  {Strodthoff}}\ and\ \bibinfo {author} {\bibfnamefont {L.}~\bibnamefont {{von
  Smekal}}},\ }\href {http://dx.doi.org/10.1016/j.physletb.2014.03.008}
  {\bibfield  {journal} {\bibinfo  {journal} {Phys.~Lett.~B}\ }\textbf
  {\bibinfo {volume} {731}},\ \bibinfo {pages} {350 } (\bibinfo {year}
  {2014})},\ \Eprint {https://arxiv.org/abs/1306.2897} {1306.2897} \BibitemShut
  {NoStop}%
\bibitem [{\citenamefont {McLerran}\ and\ \citenamefont
  {Pisarski}(2007)}]{Pisarski:0706.2191}%
  \BibitemOpen
  \bibfield  {author} {\bibinfo {author} {\bibfnamefont {L.}~\bibnamefont
  {McLerran}}\ and\ \bibinfo {author} {\bibfnamefont {R.~D.}\ \bibnamefont
  {Pisarski}},\ }\href {http://dx.doi.org/10.1016/j.nuclphysa.2007.08.013}
  {\bibfield  {journal} {\bibinfo  {journal} {Nucl.~Phys.~A}\ }\textbf
  {\bibinfo {volume} {796}},\ \bibinfo {pages} {83 } (\bibinfo {year}
  {2007})},\ \Eprint {https://arxiv.org/abs/0706.2191} {0706.2191} \BibitemShut
  {NoStop}%
\bibitem [{\citenamefont {Braguta}\ \emph {et~al.}(2016)\citenamefont
  {Braguta}, \citenamefont {Ilgenfritz}, \citenamefont {Kotov}, \citenamefont
  {Molochkov},\ and\ \citenamefont {Nikolaev}}]{Braguta:1605.04090}%
  \BibitemOpen
  \bibfield  {author} {\bibinfo {author} {\bibfnamefont {V.~V.}\ \bibnamefont
  {Braguta}}, \bibinfo {author} {\bibfnamefont {E.}~\bibnamefont {Ilgenfritz}},
  \bibinfo {author} {\bibfnamefont {A.~Y.}\ \bibnamefont {Kotov}}, \bibinfo
  {author} {\bibfnamefont {A.~V.}\ \bibnamefont {Molochkov}},\ and\ \bibinfo
  {author} {\bibfnamefont {A.~A.}\ \bibnamefont {Nikolaev}},\ }\href
  {http://dx.doi.org/10.1103/PhysRevD.94.114510} {\bibfield  {journal}
  {\bibinfo  {journal} {Phys.~Rev.~D}\ }\textbf {\bibinfo {volume} {94}},\
  \bibinfo {pages} {114510} (\bibinfo {year} {2016})},\ \Eprint
  {https://arxiv.org/abs/1605.04090} {1605.04090} \BibitemShut {NoStop}%
\bibitem [{\citenamefont {Amado}\ \emph {et~al.}(2011)\citenamefont {Amado},
  \citenamefont {Landsteiner},\ and\ \citenamefont
  {Pena-Benitez}}]{Landsteiner:1102.4577}%
  \BibitemOpen
  \bibfield  {author} {\bibinfo {author} {\bibfnamefont {I.}~\bibnamefont
  {Amado}}, \bibinfo {author} {\bibfnamefont {K.}~\bibnamefont {Landsteiner}},\
  and\ \bibinfo {author} {\bibfnamefont {F.}~\bibnamefont {Pena-Benitez}},\
  }\href {http://dx.doi.org/10.1007/JHEP05(2011)081} {\bibfield  {journal}
  {\bibinfo  {journal} {JHEP}\ }\textbf {\bibinfo {volume} {05}},\ \bibinfo
  {pages} {081}},\ \Eprint {https://arxiv.org/abs/1102.4577} {1102.4577}
  \BibitemShut {NoStop}%
\bibitem [{\citenamefont {Kaiser}\ and\ \citenamefont
  {Leutwyler}(2000)}]{Kaiser:hep-ph/0007101}%
  \BibitemOpen
  \bibfield  {author} {\bibinfo {author} {\bibfnamefont {R.}~\bibnamefont
  {Kaiser}}\ and\ \bibinfo {author} {\bibfnamefont {H.}~\bibnamefont
  {Leutwyler}},\ }\href {http://dx.doi.org/10.1007/s100520000499} {\bibfield
  {journal} {\bibinfo  {journal} {Eur.~Phys.~J.~C}\ }\textbf {\bibinfo {volume}
  {17}},\ \bibinfo {pages} {623 } (\bibinfo {year} {2000})},\ \Eprint
  {https://arxiv.org/abs/hep-ph/0007101} {hep-ph/0007101} \BibitemShut
  {NoStop}%
\bibitem [{\citenamefont {Buividovich}\ \emph {et~al.}(2021)\citenamefont
  {Buividovich}, \citenamefont {Smith},\ and\ \citenamefont {{von
  Smekal}}}]{Buividovich:21:1}%
  \BibitemOpen
  \bibfield  {author} {\bibinfo {author} {\bibfnamefont {P.~V.}\ \bibnamefont
  {Buividovich}}, \bibinfo {author} {\bibfnamefont {D.}~\bibnamefont {Smith}},\
  and\ \bibinfo {author} {\bibfnamefont {L.}~\bibnamefont {{von Smekal}}},\
  }\href@noop {} {\bibinfo {title} {Static magnetic susceptibility in
  finite-density {SU(2)} lattice gauge theory}},\ \bibinfo {howpublished}
  {Contribution to the EPJA topical issue "QCD Phase Diagram in Strong Magnetic
  Fields"} (\bibinfo {year} {2021}),\ \Eprint
  {https://arxiv.org/abs/2104.10012} {2104.10012} \BibitemShut {NoStop}%
\bibitem [{\citenamefont {Berkowitz}\ \emph {et~al.}(2017)\citenamefont
  {Berkowitz}, \citenamefont {Brantley}, \citenamefont {Bouchard},
  \citenamefont {Chang}, \citenamefont {Clark}, \citenamefont {Garron},
  \citenamefont {Joo}, \citenamefont {Kurth}, \citenamefont {Monahan},
  \citenamefont {{Monge-Camacho}}, \citenamefont {Nicholson}, \citenamefont
  {Orginos}, \citenamefont {Rinaldi}, \citenamefont {Vranas},\ and\
  \citenamefont {{Walker-Loud}}}]{Berkowitz:1704.01114}%
  \BibitemOpen
  \bibfield  {author} {\bibinfo {author} {\bibfnamefont {E.}~\bibnamefont
  {Berkowitz}}, \bibinfo {author} {\bibfnamefont {D.}~\bibnamefont {Brantley}},
  \bibinfo {author} {\bibfnamefont {C.}~\bibnamefont {Bouchard}}, \bibinfo
  {author} {\bibfnamefont {C.}~\bibnamefont {Chang}}, \bibinfo {author}
  {\bibfnamefont {M.~A.}\ \bibnamefont {Clark}}, \bibinfo {author}
  {\bibfnamefont {N.}~\bibnamefont {Garron}}, \bibinfo {author} {\bibfnamefont
  {B.}~\bibnamefont {Joo}}, \bibinfo {author} {\bibfnamefont {T.}~\bibnamefont
  {Kurth}}, \bibinfo {author} {\bibfnamefont {C.}~\bibnamefont {Monahan}},
  \bibinfo {author} {\bibfnamefont {H.}~\bibnamefont {{Monge-Camacho}}},
  \bibinfo {author} {\bibfnamefont {A.}~\bibnamefont {Nicholson}}, \bibinfo
  {author} {\bibfnamefont {K.}~\bibnamefont {Orginos}}, \bibinfo {author}
  {\bibfnamefont {E.}~\bibnamefont {Rinaldi}}, \bibinfo {author} {\bibfnamefont
  {P.}~\bibnamefont {Vranas}},\ and\ \bibinfo {author} {\bibfnamefont
  {A.}~\bibnamefont {{Walker-Loud}}},\ }\href@noop {} {\bibinfo {title} {An
  accurate calculation of the nucleon axial charge with lattice {QCD}}}
  (\bibinfo {year} {2017}),\ \Eprint {https://arxiv.org/abs/1704.01114}
  {1704.01114} \BibitemShut {NoStop}%
\bibitem [{\citenamefont {Edwards}\ \emph {et~al.}(2006)\citenamefont
  {Edwards}, \citenamefont {Fleming}, \citenamefont {Hagler}, \citenamefont
  {Negele}, \citenamefont {Orginos}, \citenamefont {Pochinsky}, \citenamefont
  {Renner}, \citenamefont {Richards},\ and\ \citenamefont
  {Schroers}}]{Edwards:hep-lat/0510062}%
  \BibitemOpen
  \bibfield  {author} {\bibinfo {author} {\bibfnamefont {R.~G.}\ \bibnamefont
  {Edwards}}, \bibinfo {author} {\bibfnamefont {G.~T.}\ \bibnamefont
  {Fleming}}, \bibinfo {author} {\bibfnamefont {P.}~\bibnamefont {Hagler}},
  \bibinfo {author} {\bibfnamefont {J.~W.}\ \bibnamefont {Negele}}, \bibinfo
  {author} {\bibfnamefont {K.}~\bibnamefont {Orginos}}, \bibinfo {author}
  {\bibfnamefont {A.}~\bibnamefont {Pochinsky}}, \bibinfo {author}
  {\bibfnamefont {D.~B.}\ \bibnamefont {Renner}}, \bibinfo {author}
  {\bibfnamefont {D.~G.}\ \bibnamefont {Richards}},\ and\ \bibinfo {author}
  {\bibfnamefont {W.}~\bibnamefont {Schroers}},\ }\href
  {http://dx.doi.org/10.1103/PhysRevLett.96.052001} {\bibfield  {journal}
  {\bibinfo  {journal} {Phys.~Rev.~Lett.}\ }\textbf {\bibinfo {volume} {96}},\
  \bibinfo {pages} {052001} (\bibinfo {year} {2006})},\ \Eprint
  {https://arxiv.org/abs/hep-lat/0510062} {hep-lat/0510062} \BibitemShut
  {NoStop}%
\bibitem [{\citenamefont {Hasenfratz}\ and\ \citenamefont
  {Knechtli}(2001)}]{Hasenfratz:hep-lat/0103029}%
  \BibitemOpen
  \bibfield  {author} {\bibinfo {author} {\bibfnamefont {A.}~\bibnamefont
  {Hasenfratz}}\ and\ \bibinfo {author} {\bibfnamefont {F.}~\bibnamefont
  {Knechtli}},\ }\href {http://dx.doi.org/10.1103/PhysRevD.64.034504}
  {\bibfield  {journal} {\bibinfo  {journal} {Phys.~Rev.~D}\ }\textbf {\bibinfo
  {volume} {64}},\ \bibinfo {pages} {034504} (\bibinfo {year} {2001})},\
  \Eprint {https://arxiv.org/abs/hep-lat/0103029} {hep-lat/0103029}
  \BibitemShut {NoStop}%
\bibitem [{\citenamefont {Bochicchio}\ \emph {et~al.}(1985)\citenamefont
  {Bochicchio}, \citenamefont {Maiani}, \citenamefont {Martinelli},
  \citenamefont {Rossi},\ and\ \citenamefont {Testa}}]{MaianiNPB262}%
  \BibitemOpen
  \bibfield  {author} {\bibinfo {author} {\bibfnamefont {M.}~\bibnamefont
  {Bochicchio}}, \bibinfo {author} {\bibfnamefont {L.}~\bibnamefont {Maiani}},
  \bibinfo {author} {\bibfnamefont {G.}~\bibnamefont {Martinelli}}, \bibinfo
  {author} {\bibfnamefont {G.~C.}\ \bibnamefont {Rossi}},\ and\ \bibinfo
  {author} {\bibfnamefont {M.}~\bibnamefont {Testa}},\ }\href
  {https://doi.org/10.1016/0550-3213(85)90290-1} {\bibfield  {journal}
  {\bibinfo  {journal} {Nucl.~Phys.~B}\ }\textbf {\bibinfo {volume} {262}},\
  \bibinfo {pages} {331} (\bibinfo {year} {1985})}\BibitemShut {NoStop}%
\bibitem [{\citenamefont {Furman}\ and\ \citenamefont
  {Shamir}(1995)}]{Furman:hep-lat/9405004}%
  \BibitemOpen
  \bibfield  {author} {\bibinfo {author} {\bibfnamefont {V.}~\bibnamefont
  {Furman}}\ and\ \bibinfo {author} {\bibfnamefont {Y.}~\bibnamefont
  {Shamir}},\ }\href {http://dx.doi.org/10.1016/0550-3213(95)00031-M}
  {\bibfield  {journal} {\bibinfo  {journal} {Nucl.~Phys.~B}\ }\textbf
  {\bibinfo {volume} {439}},\ \bibinfo {pages} {54 } (\bibinfo {year}
  {1995})},\ \Eprint {https://arxiv.org/abs/hep-lat/9405004} {hep-lat/9405004}
  \BibitemShut {NoStop}%
\bibitem [{\citenamefont {Buividovich}(2014)}]{Buividovich:13:8}%
  \BibitemOpen
  \bibfield  {author} {\bibinfo {author} {\bibfnamefont {P.~V.}\ \bibnamefont
  {Buividovich}},\ }\href {http://dx.doi.org/10.1016/j.nuclphysa.2014.02.022}
  {\bibfield  {journal} {\bibinfo  {journal} {Nucl.~Phys.~A}\ }\textbf
  {\bibinfo {volume} {925}},\ \bibinfo {pages} {218 } (\bibinfo {year}
  {2014})},\ \Eprint {https://arxiv.org/abs/1312.1843} {1312.1843} \BibitemShut
  {NoStop}%
\end{thebibliography}

%

\appendix

\section{Momentum-dependent Chiral Separation Effect for free quarks in the continuum at finite temperature}
\label{apdx:cse_free_continuum}

For free Dirac fermions, the axial-vector current-current correlator in (\ref{cse_correlator_low_momentum}) is given by the one-loop integral of the form \cite{Buividovich:13:8}:
\begin{widetext}
\begin{eqnarray}
\label{cse_free1}
 \vev{j^A_{\mu}\lr{k} j^V_{\nu}\lr{-k}} =
 T \sum\limits_{l_0} \int \frac{d^3 l}{\lr{2 \pi}^4}
 \frac{\tr\lr{\gamma_{\mu} \gamma_5 \lr{m - i \gamma_{\alpha}\lr{l_{\alpha} + k_{\alpha}/2}} \gamma_{\nu} \lr{m - i \gamma_{\beta}\lr{l_{\beta} - k_{\beta}/2}}}}{\lr{\lr{l+k/2}^2 + m^2} \lr{\lr{l - k/2}^2 + m^2}} ,
\end{eqnarray}
\end{widetext}
where $\sum\limits_{l_0}$ denotes summation over fermionic Matsubara frequencies $l_0 = 2 \pi T \lr{n + 1/2} - i \mu$ (shifted into the complex plane in order to account for the chemical potential $\mu$). We explicitly substitute the values $\mu = 1$, $\nu = 2$, $k = \lr{0, 0, 0, k_3}$ and represent the integrand as a sum of simple fractions of the form $\frac{1}{l_0 - i \mu_V \pm i \sqrt{\lr{\vec{l} \pm \vec{k}/2}^2 + m^2}}$ (see e.g. Appendix A in \cite{Buividovich:13:8} for the full derivation). The time-like momentum $l_0$ can then be summed over using the identity
\begin{eqnarray}
\label{Matsubara_sum1}
 T \sum\limits_{l_0} \frac{1}{l_0 - i \epsilon} = \frac{i}{2} \tanh\lr{\frac{\epsilon + \mu}{2 T}} .
\end{eqnarray}
After some algebraic manipulations, we obtain the following expression which is suitable for numerical integration:
\begin{widetext}
\begin{eqnarray}
\label{cse_free2}
 \vev{j^A_1\lr{k_3} j^V_2\lr{-k_3}} =
 \frac{i}{2}
 \int\limits_{-\infty}^{+\infty} \frac{d l_3}{2 \pi l_3}
 \int\limits_{0}^{+\infty} \frac{d l_{\perp}^2}{4 \pi}
 \nonumber \\
 \left(
  \tanh\lr{\frac{\mu + \sqrt{m^2 + l_{\perp}^2 + \lr{k_3/2 + l_3}^2}}{2T}}
  -
  \tanh\lr{\frac{\mu + \sqrt{m^2 + l_{\perp}^2 + \lr{k_3/2 - l_3}^2}}{2T}}
  + \right. \nonumber \\ \left. +
  \tanh\lr{\frac{\mu - \sqrt{m^2 + l_{\perp}^2 + \lr{k_3/2 + l_3}^2}}{2T}}
  -
  \tanh\lr{\frac{\mu - \sqrt{m^2 + l_{\perp}^2 + \lr{k_3/2 - l_3}^2}}{2T}}
 \right)
  ,
\end{eqnarray}
\end{widetext}

\end{document}